\documentclass[a4paper,11pt]{article}
\pdfoutput=1
\usepackage{jheppub}

\usepackage{graphicx}

\usepackage[english]{babel}
\usepackage[utf8]{inputenc}
\usepackage[T1]{fontenc}
\usepackage{amsfonts}
\usepackage{amssymb}
\usepackage{tikz}
\usepackage{nicefrac}
\usepackage{graphicx}
\usepackage{hyperref}
\usepackage{soul}

\usetikzlibrary{decorations.pathmorphing}
\usetikzlibrary{decorations.pathreplacing}
\usetikzlibrary{decorations.markings}
\usetikzlibrary{arrows}
\usepackage{multirow}
\usepackage{bbold}
 
\newcommand{\be}{\begin{equation}}
\newcommand{\ee}{\end{equation}}
\newcommand{\bea}{\begin{eqnarray}}
\newcommand{\eea}{\end{eqnarray}}

\newcommand{\Tr}{\textrm{Tr}}

\newcommand{\bra}{\langle}
\newcommand{\ket}{\rangle}

\newcommand{\eps}{\epsilon}

\newcommand{\cO}{\mathcal{O}}

\newcommand{\half}{\frac{1}{2}}

\begin{document}
\title{The QCD phase diagram in the limit of heavy quarks using complex
Langevin dynamics}

\author[a]{Gert Aarts,}
\author[a]{Felipe Attanasio,}
\author[a]{Benjamin J\"{a}ger}
\author[b,c]{and D\'{e}nes Sexty}
\affiliation[a]{Department of Physics, College of Science, Swansea University,\\
Swansea, SA2 8PP, United Kingdom}
\affiliation[b]{Department of Physics, Bergische Universit\"{a}t Wuppertal,\\
Gaussstra{\ss}e 20, D-42119 Wuppertal, Germany}
\affiliation[c]{Inst. for Theoretical Physics, E\"otv\"os University,\\
P\'azm\'any P. s\'et\'any 1/A, H-1117 Budapest, Hungary}

\emailAdd{g.aarts@swan.ac.uk}
\emailAdd{pyfelipe@swan.ac.uk}
\emailAdd{b.jaeger@swan.ac.uk}
\emailAdd{sexty@uni-wuppertal.de}

\abstract{Complex Langevin simulations allow numerical studies of theories that exhibit
		a sign problem, such as QCD, and are thereby potentially suitable to determine the QCD
		phase diagram from first principles. Here we study QCD in the limit of heavy
		quarks for a wide range of temperatures and chemical potentials. Our
		results include an analysis of the adaptive gauge cooling technique, which
		prevents large excursions into the non-compact directions of
		the SL($3, \mathbb{C}$) manifold. We find that such excursions may appear
		spontaneously and change the statistical distribution of physical
		observables, which leads to disagreement with known results.
		Results whose excursions are sufficiently small are used to map the
		boundary line between confined and deconfined quark phases.}

\date{\today}


\arxivnumber{1606.05561}

\maketitle

\section{Introduction}
Strongly-interacting matter at nonzero temperature and baryon density, in both the hadronic phase and the quark-gluon plasma, has been the subject of active research. Experimentally, it can be investigated by colliding heavy ions, and this programme is running successfully at the Relativistic Heavy Ion Collider (BNL) and the Large Hadron Collider (CERN). 
On the theoretical side, nonperturbative studies of the theory of the strong interaction, Quantum
Chromodynamics, at finite temperature have nowadays reached maturity, by employing the lattice discretisation \cite{Borsanyi:2013bia,Bazavov:2014pvz}.
However, at nonzero density (or baryon chemical potential) numerical lattice simulations have to overcome the {\em sign problem}, since the Boltzmann weight in the partition function is complex. This severely limits the applicability of standard numerical approaches~\cite{deForcrand:2010ys}.
As a consequence, many alternative numerical lattice field theory approaches
have been proposed and recent reviews can be found in
Refs.~\cite{Aarts:2013lcm,Sexty:2014dxa,Scorzato:2015qts,Gattringer:2016kco}.

In this paper we use the complex Langevin (CL) method to study QCD at finite temperature and
chemical potential in the limit of heavy quarks (heavy dense QCD, HDQCD).
This model shares many features with QCD with fully dynamical quarks which are
interesting from a numerical point of view, such as the {\em sign} and {\em Silver
Blaze} \cite{Cohen:2003kd} problems, but is considerably cheaper in terms of computer time.
Indeed, this limit of QCD \cite{Bender:1992gn} has been studied using a variety of approaches, e.g.\ in combination with a strong-coupling expansion~\cite{deForcrand:2014tha,Glesaaen:2015vtp}, reweighting \cite{DePietri:2007ak}, and by employing a histogram \cite{Ejiri:2013lia} or density of states method \cite{Garron:2016noc}. Previous CL studies \cite{Aarts:2008rr,Aarts:2009dg,Seiler:2012wz,Aarts:2014bwa} have focussed mostly on the method, leading to important algorithmic improvements. Further discussion of HDQCD can be found in Refs.~\cite{Rindlisbacher:2015pea,Seiler:2015uwe}.
We emphasise that the CL method allows for direct simulations throughout the phase diagram, without the need for further approximation or reweighting. In particular, in contrast to strong-coupling approaches, the gluonic dynamics is contained without approximation and hence e.g.\ the thermal deconfinement transition at $\mu = 0$ is captured correctly (in the presence of heavy quarks).
We therefore have the opportunity to map the phase boundary, connecting the
thermal transition at high temperature with the onset transition, where the
quark density becomes non-zero, at large chemical potential. As such, it is a good test scenario to prepare for the realistic case of QCD with light quarks.

The CL method consists of stochastic explorations of a complexified
configuration space, without the requirement of a positive
weight~\cite{Parisi:1984cs, Klauder:1983nn, Klauder:1983zm, Klauder:1983sp}.
It is precisely the method's complex nature that allows for a solution of the
sign problem, even when it is severe~\cite{Aarts:2008rr, Aarts:2008wh,
Aarts:2010gr, Aarts:2011zn}. However, success is not
guaranteed~\cite{Ambjorn:1985iw, Ambjorn:1986fz, Berges:2006xc, Berges:2007nr, Aarts:2010aq, Pawlowski:2013pje, Pawlowski:2013gag} and convergence to a wrong limit may occur. Based on the theoretical justification of the approach \cite{Aarts:2009uq,Aarts:2011ax}, these cases of incorrect convergence can be identified a posteriori.
Here we employ the adaptive gauge cooling technique~\cite{Seiler:2012wz,
Aarts:2013uxa}, which is necessary but not sufficient to avoid convergence to wrong limits.
In addition, in the presence of a fermion determinant, the drift appearing in the CL equation is no longer holomorphic, which requires a reconsideration of the justification \cite{Aarts:2009uq,Aarts:2011ax} and may lead again to incorrect convergence in practice \cite{Mollgaard:2013qra,Splittorff:2014zca,Nishimura:2015pba}. However, all indications are that this is not an issue for the model considered in this paper \cite{Seiler:2012wz,Aarts:2014bwa}.
We remark that applications of CL to full QCD can be found in Refs.~\cite{Sexty:2013ica,Aarts:2014bwa} and a comparison with multi-parameter reweighting in Ref.~\cite{Fodor:2015doa}.

This paper proceeds as follows: In section \ref{sec.cle.gc} we review the
complex Langevin method applied to lattice QCD and the gauge cooling
technique. Section~\ref{sec.HDQCD.details} presents the heavy dense (HD)
approximation of QCD and lists the parameters used in our simulations.
In section~\ref{sec.results} we present results for observables related to the Polyakov loop and quark density as functions of the temperature and chemical potential as well as the resulting phase diagram.
Issues related to instabilities and their relation to excursion into the non-compact directions of the
Langevin equations are discussed in section~\ref{sec.insta}.
In section~\ref{sec.concl} we present a conclusion and an outlook for future work.
 Preliminary results have appeared in Refs.~\cite{Aarts:2014kja, Aarts:2014fsa,
Aarts:2015yba, Aarts:2015yuz}.

\section{\label{sec.cle.gc}Complex Langevin equation and gauge cooling}

We consider QCD in the grand-canonical formulation, where the (quark) chemical potential $\mu$ couples to quark number. For an elementary introduction, see e.g.\ Ref.\ \cite{Aarts:2015tyj}. After integrating out the bilinear quark fields, the partition function is written as
\be
Z= \int DU\, e^{-S_{\rm YM}} \det M \equiv \int DU\, e^{-S}, 
\qquad\qquad
S = S_{\rm YM} - \ln\det M,
\ee
where $S_{\rm YM}$ is the Yang-Mills action, $U$ are the gauge links, and $M$ is the fermion matrix, depending on the chemical potential and the gauge links. 
 Quantum expectation values can be evaluated using Langevin dynamics in a procedure
known as stochastic quantisation~\cite{Parisi:1980ys}. In this scheme,
expectation values are obtained as averages over a stochastic process by evolving dynamical 
variables over a fictitious time $\theta$. Importantly, importance sampling does not enter in this formulation.

On the lattice, for an SU($3$) gauge theory with links $U_{x,\nu}$, a Langevin update, using a
first-order discretisation in the Langevin time $\theta=n\eps$, reads~\cite{Damgaard:1987rr}
\begin{equation}
	U_{x,\nu}(\theta + \varepsilon) = \exp \left[ i \lambda^a \left( 
	\varepsilon K^a_{x,\nu} + \sqrt{\varepsilon} \, \eta^a_{x,\nu} \right) \right] U_{x,\nu}(\theta),
	\label{eq.Langevin}
\end{equation}
where $\lambda^a$ are the Gell-Mann matrices (with $\Tr\, \lambda^a\lambda^b=2\delta^{ab}$, the sum over $a=1,\ldots, 8$ is assumed) and $\eta^a_{x,\nu}$ are Gaussian white noise fields, which satisfy 
\begin{equation}
\left< \eta^a_{x,\mu}
\,\eta^b_{y,\nu} \right> =
2\,\delta_{xy}\,\delta^{ab}\,\delta_{\mu\nu}.
\end{equation}
The dynamics is governed by the action $S$, which generates the drift 
\begin{equation}
K^a_{x,\nu} = -D^a_{x,\nu} S = -D^a_{x,\nu} S_{\rm YM} + \Tr\left[ M^{-1} D^a_{x,\nu} M \right],
\end{equation}
where $D^a_{x,\nu}$ is the
gauge group derivative
\begin{equation}
D^a_{x,\nu} \,f(U) = \frac{\partial}{\partial \alpha} f \big(
\mathrm{e}^{i\,\alpha\, \lambda_a } \,U_{x,\nu} \big) \Big|_{\alpha = 0} .
\end{equation}
The quark contribution leads to poles in the drift, namely where $\det M=0$ and $M^{-1}$ does not exist. In some cases this affects the results negatively \cite{Mollgaard:2013qra,Nishimura:2015pba}, but in HDQCD this is not the case, as far as is understood~\cite{Splittorff:2014zca,Aarts:2014bwa}.
In order to avoid numerical instabilities and regulate large values of the drift, it is necessary to change the
Langevin stepsize $\varepsilon$ adaptively~\cite{Aarts:2009dg},
based on the absolute value of the drift term $K^a_{x,\nu}$.

In theories that exhibit the sign problem the drift is complex, resulting in an exploration of a larger configuration space. This is how the sign problem is potentially evaded~\cite{Klauder:1983nn, Klauder:1983zm, Klauder:1983sp,
Parisi:1984cs, Aarts:2008rr,Aarts:2009uq}.
In an SU($3$) gauge theory, this procedure enlarges the gauge group to SL($3,\mathbb{C}$). 
The latter group, however, is not compact.
Parametrising the gauge links as 
\be
 U_{x,\nu} = \exp \left[ i\lambda^a A^a_{x,\nu} \right],
 \ee
 this implies that the gauge fields $A^a_{x,\nu}$ can now assume complex values. The extra degrees
of freedom can lead to trajectories in which the imaginary parts of 
the gauge fields are not a small deformation. A measure of the distance from the
unitary manifold can be given by unitarity norms 
\be
\label{eq:norm}
	d_1 = \frac{1}{3 \Omega} \sum_{x, \nu} \Tr\left[ U_{x,\nu} U_{x,\nu}^\dagger - \mathbb{1} \right] \geq 0, 
	\qquad
	d_2 = \frac{1}{3 \Omega} \sum_{x, \nu} \Tr\left[ U_{x,\nu} U_{x,\nu}^\dagger - \mathbb{1} \right]^2 \geq 0,
\ee
etc., 
where $\Omega=N_\tau N_s^3$ is the four dimensional simulation volume.
These norms are invariant under unitary gauge transformations, but not under general SL($3,\mathbb{C}$) transformations. They are exactly zero only if all links $U_{x,\nu}$ are unitary.

Gauge cooling~\cite{Seiler:2012wz} is a procedure to reduce the distance to
the unitary manifold via SL($3,\mathbb{C}$) gauge transformations. It consists of a sequence of gauge
transformations which decrease the unitary norms $d_i$ in a steepest descent
fashion
\begin{equation}
		U_{x,\nu} \to e^{-\varepsilon \alpha \lambda^a f^a_x} \, U_{x,\nu} \, 
		e^{\varepsilon \alpha \lambda^a f^a_x}, 
		\qquad 
		f^a_x = 2 \sum_\nu\Tr \left[ \lambda^a
		\left( U_{x,\nu} U^\dagger_{x,\nu} - U^\dagger_{x-\nu,\nu} U_{x-\nu,\nu}
		\right) \right].
\end{equation}
Note that $f^a_x$ is obtained via an infinitesimal gauge transformation of $d_1$.
In order to optimise the cooling procedure, the coefficient $\alpha$ is changed adaptively based on the absolute value of
$f^a_x$~\cite{Aarts:2013uxa}. Cooling is also stopped once the rate of change of the unitary norm is below a set target.

\section{\label{sec.HDQCD.details}Heavy dense QCD}

We consider the heavy dense approximation of QCD (HDQCD)~\cite{Bender:1992gn, Aarts:2008rr}, in which the gluonic action is the standard Wilson Yang-Mills lattice action, while in the quark action spatial hopping terms are neglected but all chemical potential dependence, which resides in the temporal hopping terms, is retained. As mentioned earlier, the gluonic dynamics is contained without approximation.

The action $S$ then consists of the gluonic term,
\begin{equation}
		S_{\rm YM} = -\frac{\beta}{6} \sum_x \sum_{\mu < \nu} \Tr \left[ 
		U_{x,\mu\nu} + U^{-1}_{x,\mu\nu} \right],
\end{equation}
 where $U_{x,\mu\nu} = U_{x,\mu} U_{x+\mu,\nu} U^{-1}_{x+\nu,\mu} U^{-1}_{x,\nu}$
is the standard plaquette and $\beta$ the lattice gauge coupling, and minus the logarithm of the quark determinant in the HD approximation.
The latter is obtained from the standard Wilson fermion action,
\begin{equation}
	M_{xy} = \delta_{x,y} - 2\kappa \sum_{\nu=1}^4 \left( e^{\delta_{\nu,4} \mu} \Gamma_{-\nu} U_{x,\nu} \delta_{x+a\nu, y} + e^{-\delta_{\nu,4} \mu} \Gamma_{+\nu} U^{-1}_{x,\nu} \delta_{x-a\nu, y} \right)\,,
\end{equation}
by dropping the spatial hopping terms, such that
\begin{equation}
	M_{xy} = \delta_{x,y} - 2\kappa \left( e^{\mu} \Gamma_{-4} U_{x,4} \delta_{x+a \hat{4}, y} + e^{-\mu} \Gamma_{+4} U^{-1}_{x,4} \delta_{x-a\hat{4}, y} \right)\,,
\end{equation}
where $\Gamma_{\pm \nu} = (\mathbf{1} \pm \gamma_\nu)/2$.
Taking the determinant in Dirac space and in spacetime indices yields, for a single quark flavour (below we consider $N_f = 2$ degenerate quarks),
\begin{equation}
 \label{eq:det}
		\det M = \prod_{N_f} \prod_{\vec{x}} 
		\left\{ 
		\det\left[ 1 + h e^{\mu/T} \mathcal{P}_{\vec{x}} \right]^2 
		\det\left[ 1 + h e^{-\mu/T} \mathcal{P}^{-1}_{\vec{x}} \right]^2
		\right\}.
\end{equation}
The power $2$ originates from the gamma-matrix structure and the $+$ sign from the anti-periodic boundary conditions.
In this expression $\mathcal{P}_{\vec{x}}^{(-1)}$ are the (inverse) Polyakov loops, 
\begin{equation}
		\mathcal{P}_{\vec{x}} = \prod_{\tau = 0}^{N_\tau - 1} U_{(\vec{x},\tau),4}		
		\qquad \mathrm{and} \qquad 
		\mathcal{P}^{-1}_{\vec{x}} = \prod^0_{\tau=N_\tau-1} U^{-1}_{(\vec{x},\tau), 4},
\end{equation}
with $N_\tau$ the number of time slices in the temporal direction. The temperature $T$ is related to $N_\tau$ via $T=1/(a N_\tau)$, with $a$ the lattice spacing. The parameter $h=(2\kappa)^{N_{\tau}}$, with $\kappa$ the hopping parameter, arises from the hopping expansion and, finally, $N_f$ is the number of quark flavours.

Important observables are the expectation value of the traced (inverse) Polyakov loops and the quark density, defined by 
\begin{align}
		&\bra P\ket = \frac{1}{V} \sum_{\vec{x}} \bra P_{\vec{x}} \ket, 
		\qquad\qquad\qquad\qquad
		 P_{\vec{x}} = \frac{1}{3} \Tr\, \mathcal{P}_{\vec{x}},
		\\
		&\bra P^{-1}\ket = \frac{1}{V} \sum_{\vec{x}} \bra P^{-1}_{\vec{x}}\ket, 
		\qquad\qquad\qquad
		 P^{-1}_{\vec{x}} = \frac{1}{3} \Tr\, \mathcal{P}^{-1}_{\vec{x}},
		\\
		& \langle n \rangle = \frac{T}{V} \frac{\partial \ln Z}{\partial \mu} = \frac{1}{V} \sum_{\vec{x}} \bra n_{\vec{x}}\ket, 
\end{align}
 with $V=N_s^3$ the spatial volume, and \cite{Aarts:2008rr}
 \be
 n_{\vec{x}} = 6N_f \frac{ zP_{\vec{x}} +2 z^2 P_{\vec{x}}^{-1} +z^3}{ 1+ 3zP_{\vec{x}} + 3 z^2 P_{\vec{x}}^{-1} +z^3} 
 -6N_f \frac{ \bar z P_{\vec{x}}^{-1} +2 \bar z^2 P_{\vec{x}} +\bar z^3}{ 1+ 3\bar zP_{\vec{x}}^{-1} + 3 \bar z^2 P_{\vec{x}} +\bar z^3}.
 \ee
Here we used the notation
\be 
 z = h e^{\mu/T}, \qquad\qquad \bar z = h e^{-\mu/T}.
 \ee
We note here that $P_{\vec{x}}$ and $P^{-1}_{\vec{x}}$ are complex-valued for a given gauge configuration but that their
expectation values are real, as they are related to the free energy of a
single (anti) quark. Below we will also consider the symmetrised combination
\be
\label{eq:Ps}
 P^{\rm s}_{\vec{x}} = \half\left( P+P^{-1}\right),
\ee
which is real for each SU(3) gauge link configuration.

It is useful to consider the zero-temperature limit, $N_\tau\to \infty$, at fixed lattice spacing.
We take $\mu>0$ and first look at the density. The contribution from the
anti-quarks, i.e.\ the second term in Eq.~\ref{eq:norm}, is exponentially
suppressed. For the quark contribution, we write $z$ as
\be
 z = h e^{\mu/T} = \left(2\kappa e^\mu\right)^{N_\tau} \equiv \exp \left[ \left(\mu-\mu_c^0\right) N_\tau\right], \qquad\qquad \mu_c^0 \equiv -\ln\left(2\kappa\right),
\ee 
where we used that $\mu/T = \mu N_\tau$, with $\mu$ expressed in lattice units after the equality sign.
We see therefore that at zero temperature the density vanishes when $\mu<\mu_c^0$ (Silver Blaze region \cite{Cohen:2003kd,Aarts:2015tyj}) and equals saturation density ($n_{\rm sat} = 6N_f$) when $\mu>\mu_c^0$, irrespective of the value of the Polyakov loop. 
Hence $\mu_c^0$ is the critical chemical potential for onset at $T=0$, but the behaviour in the region $\mu>\mu_c^0$ is a lattice artefact. For the Polyakov loop, we similarly note that at zero temperature and $\mu<\mu_c^0$, the quarks do not couple to the gauge fields and hence $\bra P\ket=0$, as in the pure gauge theory, while when $\mu>\mu_c^0$, $\bra P\ket$ has to be zero as well to ensure a finite determinant. Hence at $T=0$, $\bra P\ket=0$ for all $\mu$, except possibly at $\mu=\mu_c^0$. The vanishing of $\bra P\ket$ above onset is again due to the maximal number of quarks that can be placed on a finite lattice. 
For more discussion of these aspects, see e.g.\ Ref.\ \cite{Rindlisbacher:2015pea}. 

\begin{table}[h]
		\begin{tabular}{ccccccccccccccc}
				\hline
				\multicolumn{5}{c|}{$\beta = 5.8$} & \multicolumn{5}{c|}{$V = 6^3, 8^3,
				10^3$} & \multicolumn{5}{c}{$a\sim 0.15$ fm} \\
				\multicolumn{5}{c|}{$\kappa = 0.04$} & \multicolumn{5}{c|}{$N_f = 2$} &
				\multicolumn{5}{c}{$\mu_c^0=2.53$} \\
				\hline
				$N_\tau$ & 28 & 24 & 20 & 16 & 14 & 12 & 10 & 8 & 7 & 6 & 5 & 4 & 3 & 2 \\
				$T$ [MeV] & 48 & 56 & 67 & 84 & 96 & 112 & 134 & 168 & 192 & 224 & 268 &
				336 & 447 & 671\\  
				\hline
		\end{tabular}
		\caption{\label{tb.simulation.parameters}  
		Parameters used in this study. The chemical potential $\mu$ is varied from 0 to $1.3\mu_c^0$, with 
		$\mu_c^0=-\ln(2\kappa)$. The lattice spacing is set using the gradient flow~\cite{Borsanyi:2012zs} and 
		is approximate.		
		}
\end{table}

Simulation parameters are listed in Table~\ref{tb.simulation.parameters}. In order to scan the phase diagram, a wide range of temperatures and chemical potentials is covered and a total of $880$ ensembles with different combinations of $N_\tau$ and $\mu$ were generated, for each of the three volumes. We use a fixed
gauge coupling throughout this work, $\beta=5.8$, and the estimate of the lattice spacing of $a \sim 0.15\text{ fm}$ has been obtained using the gradient flow~\cite{Borsanyi:2012zs}. Using a fixed lattice spacing yields an adequate coverage of the phase diagram at low temperature, with fixed lattice artefacts, but a poorer coverage at larger temperature.

\section{\label{sec.results}Phase diagram}

We have performed an extensive scan of $T-\mu$ plane, to determine the phase structure by direct simulation \cite{link}. 
In order to track the reliability \cite{Aarts:2009uq,Aarts:2011ax} of the results we measured the unitarity norms (\ref{eq:norm}), studied the distributions of observables, and compared with results obtained with reweighting \cite{DePietri:2007ak}, where applicable.
From this analysis, we inferred that complex Langevin dynamics in combination with gauge cooling produces correct results, provided that the unitarity norm does not become too large, $d_2 \lesssim \cO(0.1)$. In light of these observations we present here only simulation data for which the unitarity norm is smaller than $0.03$. In this regime we can
extract physical information on the phase boundary of HDQCD. We come back to larger unitarity norms in Sec.\ \ref{sec.insta}.

\begin{figure}[!ht]
	\begin{tabular}{cc}
	\includegraphics[scale=0.46,clip, trim=0.5cm 0.5cm 0cm
	0cm]{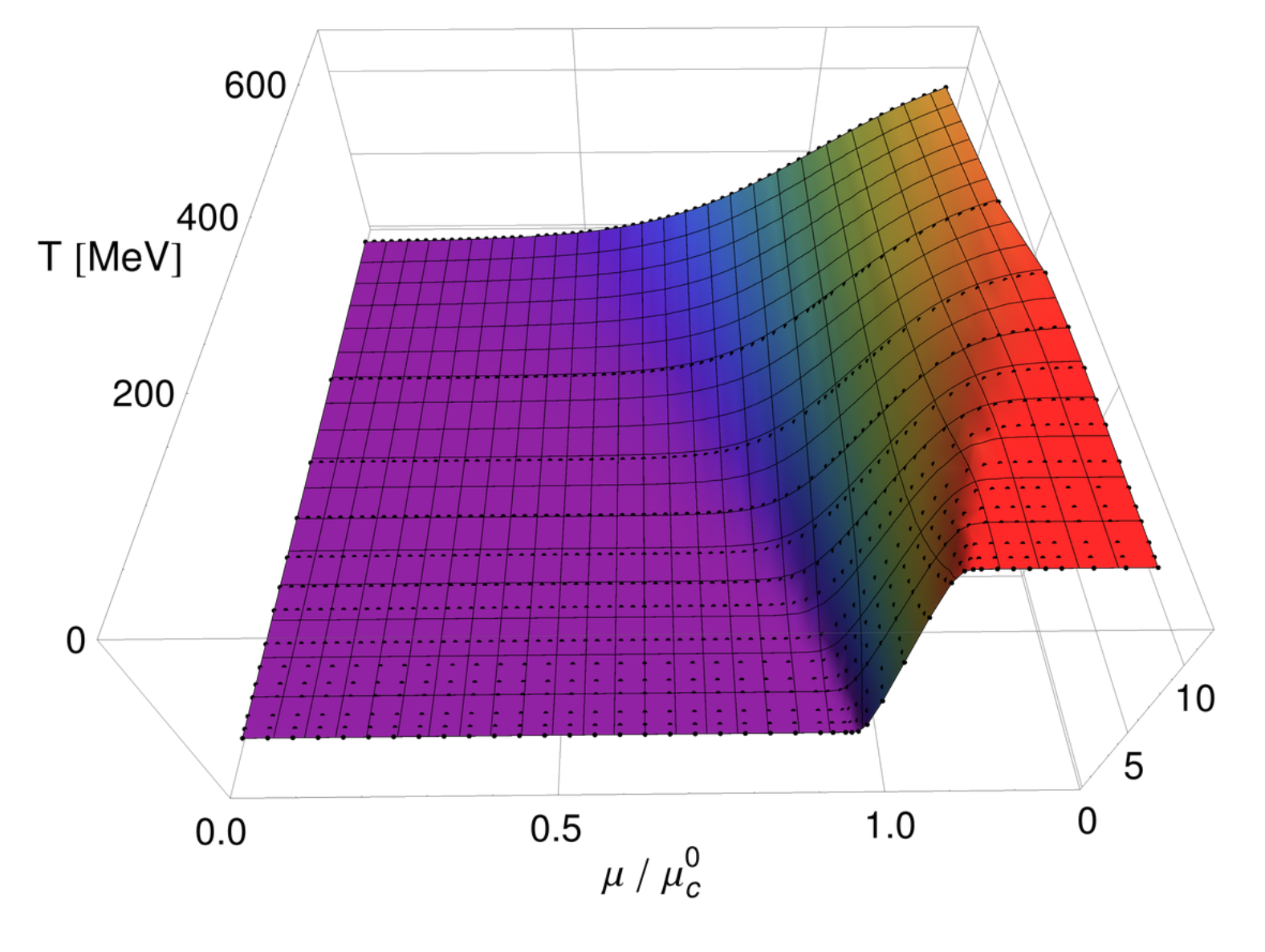} &
	\includegraphics[scale=0.46,clip, trim=0.5cm 0.5cm 0cm
	0cm]{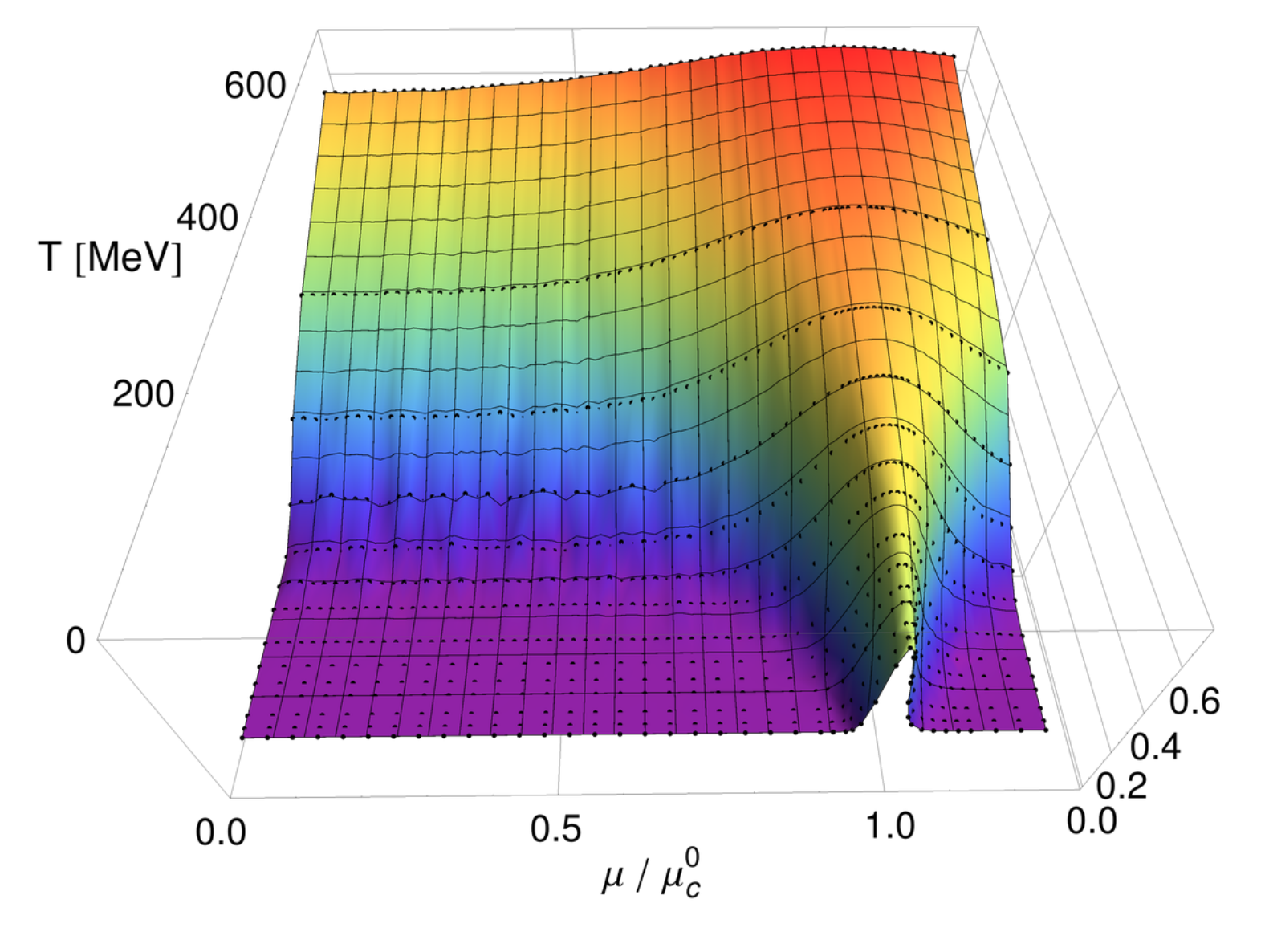}
	\end{tabular}
	\caption{\label{fig.observable.3d} Quark density $\bra n\ket$ (left) and 
	symmetrised Polyakov loop $\half\bra P+P^{-1}\ket$ (right) as functions of the temperature and chemical potential on a $10^3$ volume. The black points correspond to the simulations' results. The coloured surface is a cubic spline to guide the eye.
	}
\end{figure}

Figure~\ref{fig.observable.3d} shows the quark density $\bra n\ket$ and the symmetrised Polyakov loop $\bra P^{\rm s}\ket = \half\bra P+P^{-1}\ket$ as functions of the temperature and chemical potential on the spatial volume of $10^3$.
The plotted surfaces are cubic splines to guide the eye and each black point represents the average from an individual simulation.
Other parameters are given in Table \ref{tb.simulation.parameters}. We have used the lattice spacing of $a \sim 0.15$ fm to convert the temperature
to physical units and expressed the chemical potential in terms of $\mu^0_c$. The
Polyakov loop shows both the thermal deconfinement transition, driven by gluonic dynamics, and the transition to high
densities, driven by quark dynamics. The region where $\mu>\mu_c^0$ is a lattice artefact and the Polyakov loop drops again to zero at low temperature, as explained above. At higher temperature, the Polyakov loop is nonzero for all chemical potentials.
At low temperature the quark density rises sharply at $\mu=\mu_c^0$ to saturation density ($n_{\rm sat}=12$). This behaviour is smoothened out at higher temperature. The density only rises slowly as $\mu$ increases from zero; for heavy quarks, the quark number susceptibility at $\mu=0$ is exponentially suppressed.

\begin{figure}[!ht]
	\begin{tabular}{cc}
	\includegraphics[scale=0.46,clip, trim=0.5cm 0.5cm 0cm
	0cm]{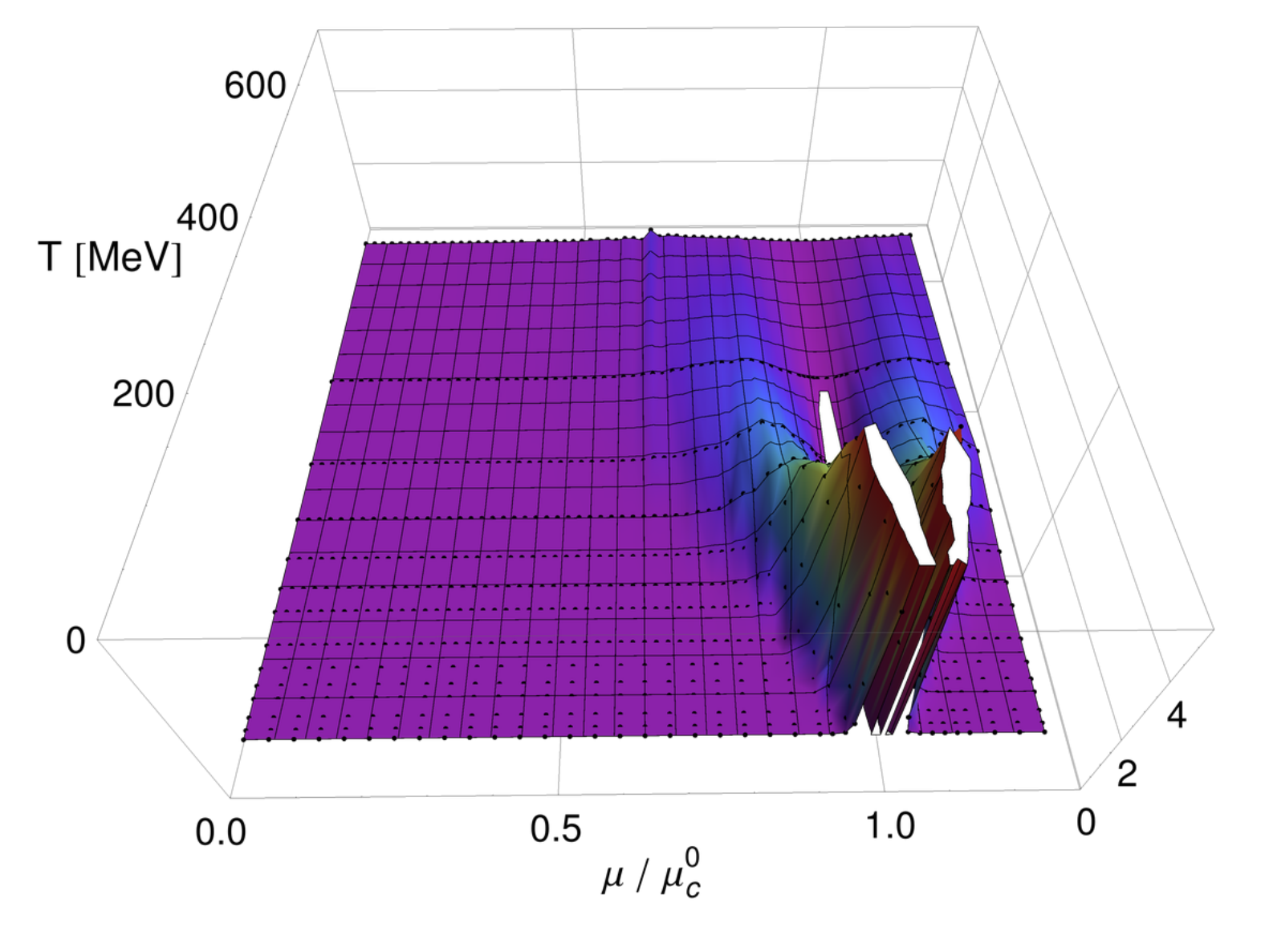} &
	\includegraphics[scale=0.46,clip, trim=0.5cm 0.5cm 0cm
	0cm]{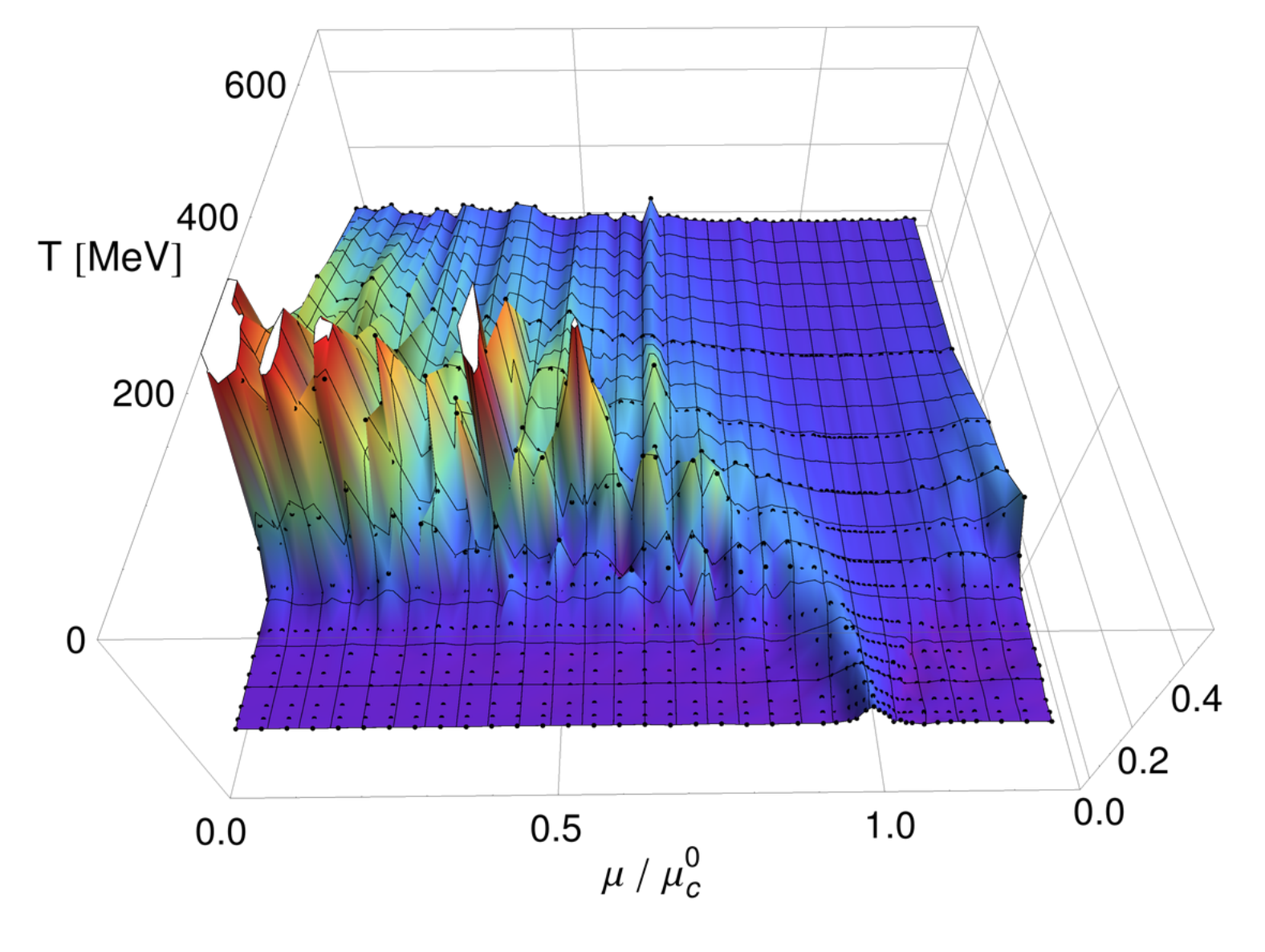}
	\end{tabular}
	\caption{\label{fig.susceptibility.3d} Susceptibility of the quark density $\bra
	n\ket$ (left) and symmetrised Polyakov loop $\half\bra P+P^{-1}\ket$ (right).
	In both cases peak heights have been cut, resulting in white
	plateaus.}
\end{figure}

Figure~\ref{fig.susceptibility.3d} shows the susceptibilities for the
aforementioned observables, which outline the corresponding transitions.
Note that in both cases dominant peaks are not shown, to improve
visibility.
In principle the phase boundary can be determined from these
susceptibilities. A better signal, however, is obtained by employing the Binder
cumulant $B$ \cite{Binder:1981sa}, which for an observable $O$ is defined as
\be
 B = 1 - \frac{\bra O^4\ket}{3\bra O^2\ket^2}.
\ee
Let $\bra O\ket$ be zero in one phase and nonzero in another, and assume that the higher moments are governed by Gaussian fluctuations. It is then easy to see that
\be
\bra O\ket = 0 \Leftrightarrow B=0, 
\qquad\qquad
\bra O\ket \neq 0 \Leftrightarrow B=\frac{2}{3},
\ee
where in the latter case it is assumed that $\bra O^2\ket - \bra O\ket^2 \ll \bra O\ket^2$.

\begin{figure}[t]
	\begin{tabular}{cc}
	\includegraphics[scale=0.50,clip, trim=0.5cm 0.5cm 0cm
	0cm]{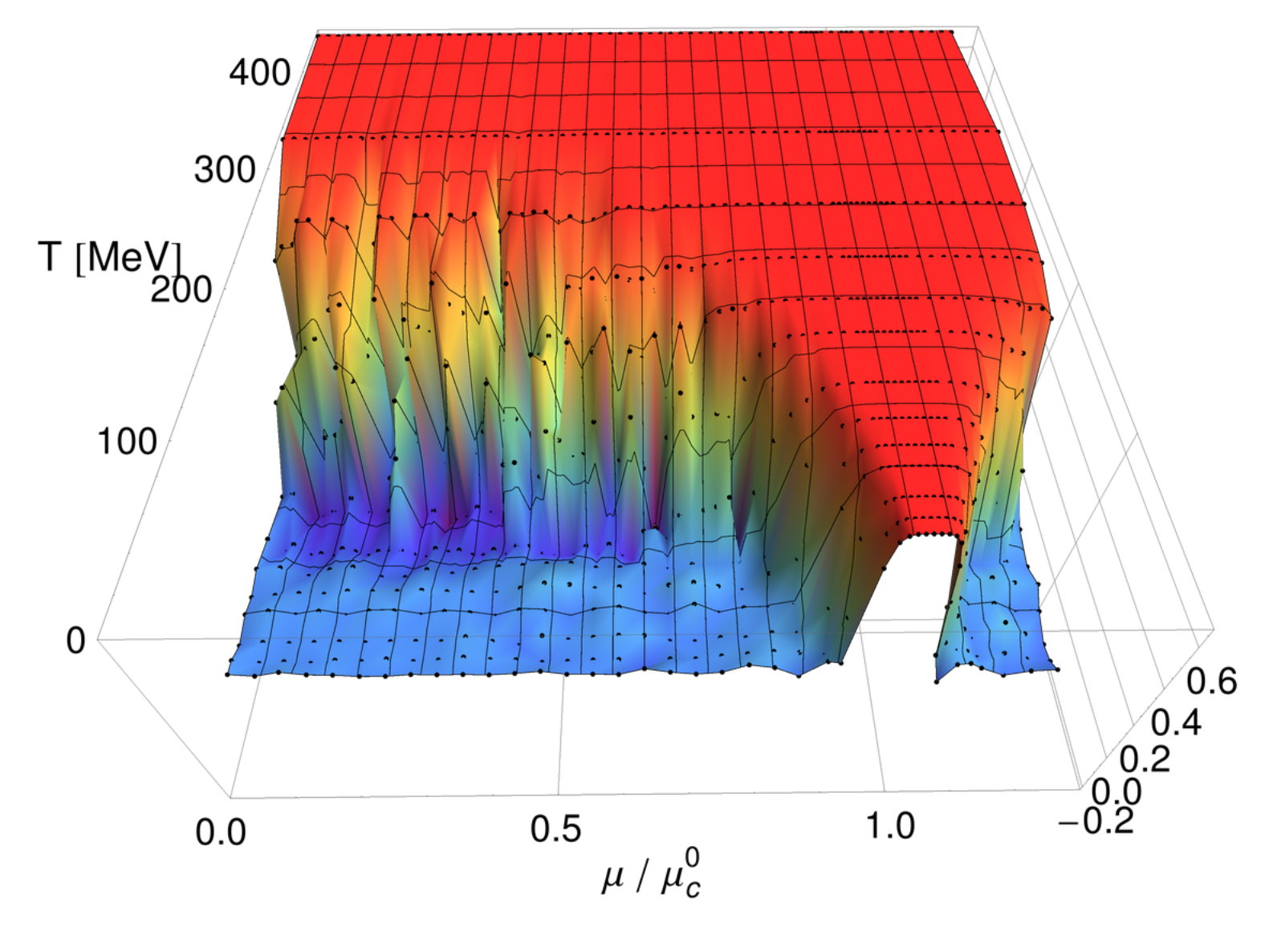} 
	\includegraphics[scale=0.85, clip, trim=3cm 0cm 0cm
	0cm,scale=0.30]{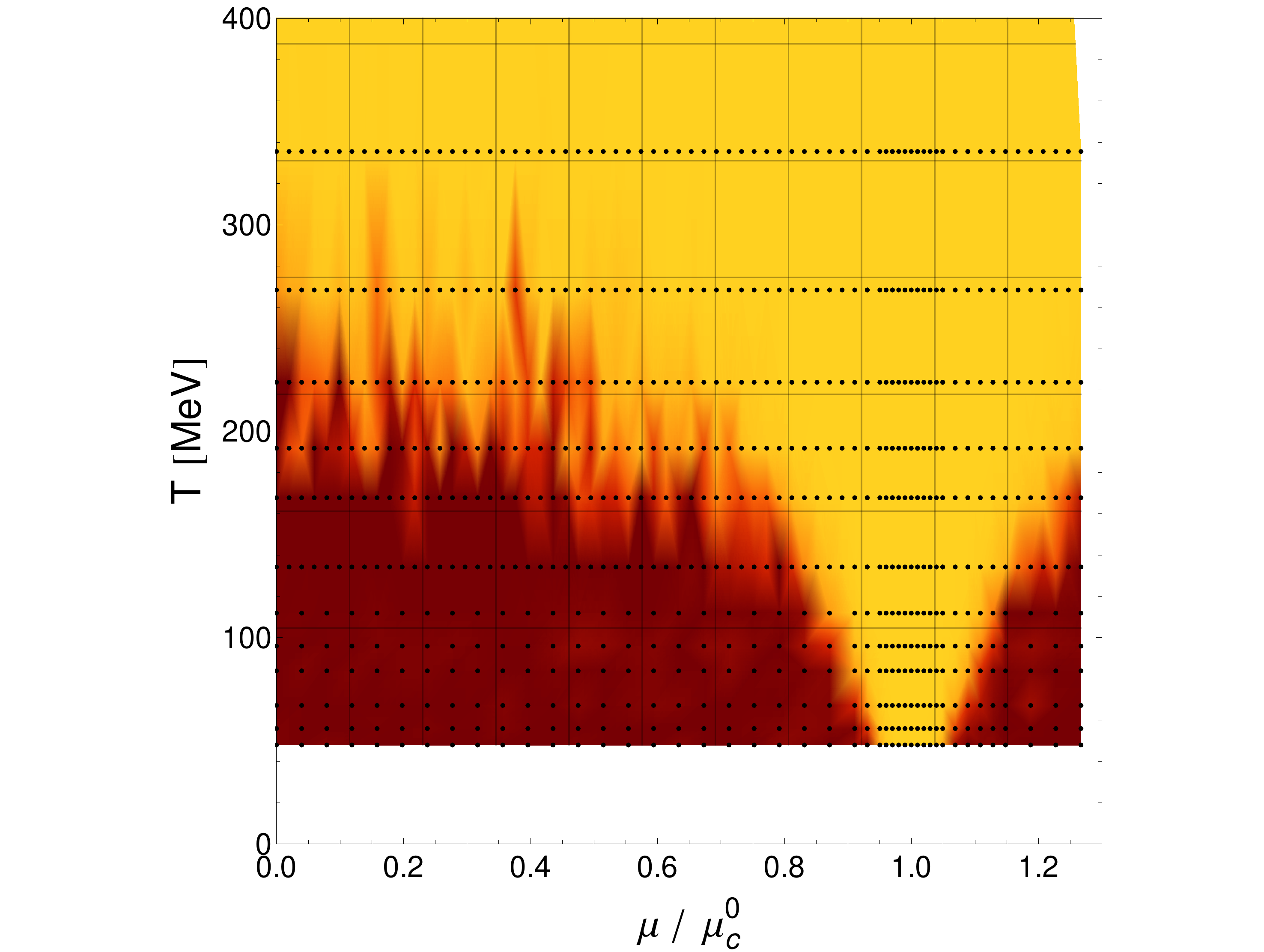}
	\end{tabular}
	\caption{\label{fig.Binder} Left: Binder cumulant of the symmetrised Polyakov loop
	as function of $T$ and $\mu$. Right: Two dimensional projection of the
	Binder cumulant. Red colours indicate a value compatible
	with 0, whereas yellow shows the region for which the Binder cumulant is 2/3.
	}
\end{figure}

\begin{figure}[t] 
	\begin{tabular}{cc}
	\includegraphics[scale=0.60,clip, trim=0.1cm 0cm 0cm 0cm]{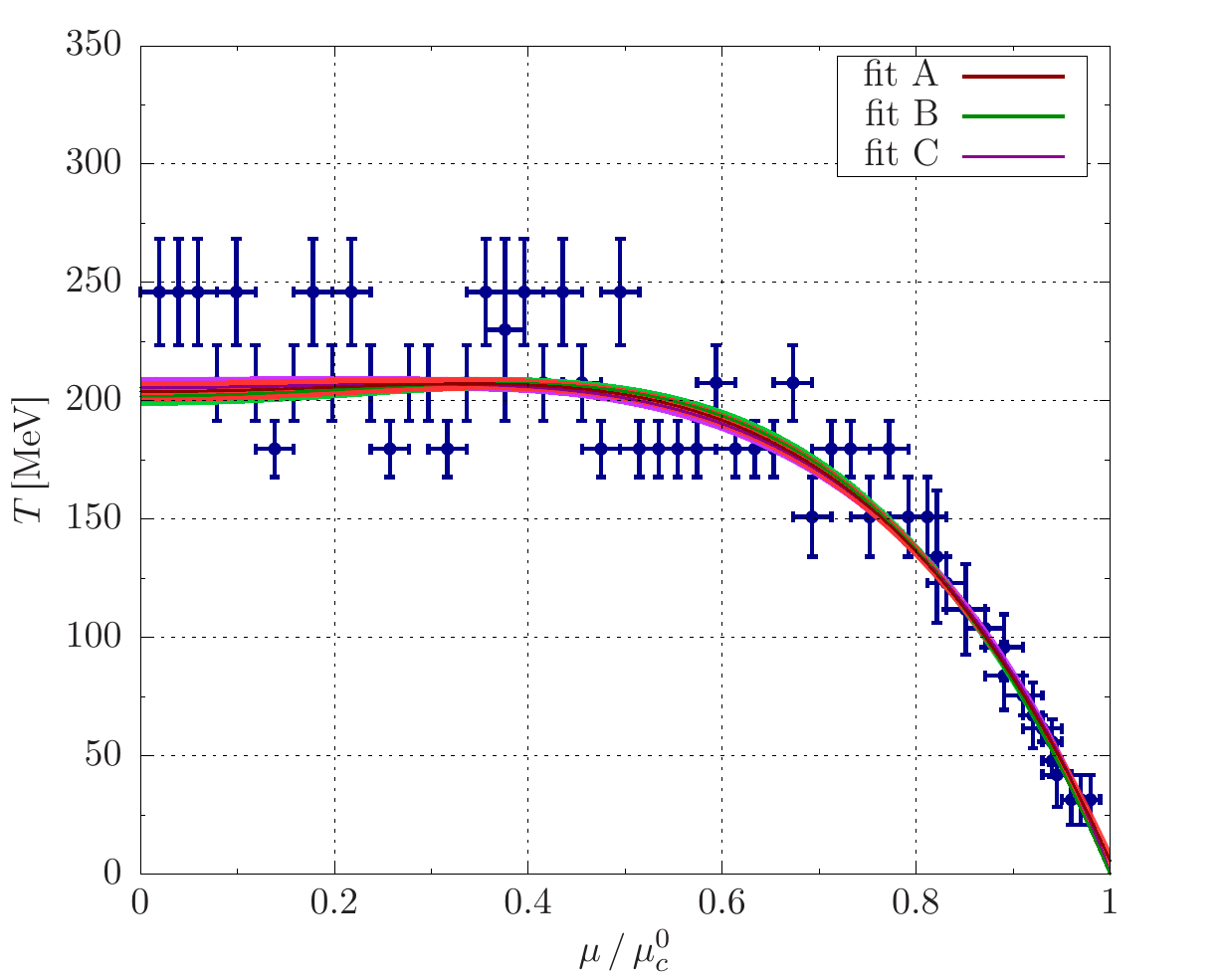}
	&
	\includegraphics[scale=0.60,clip, trim=0.1cm 0cm 0cm 0cm]{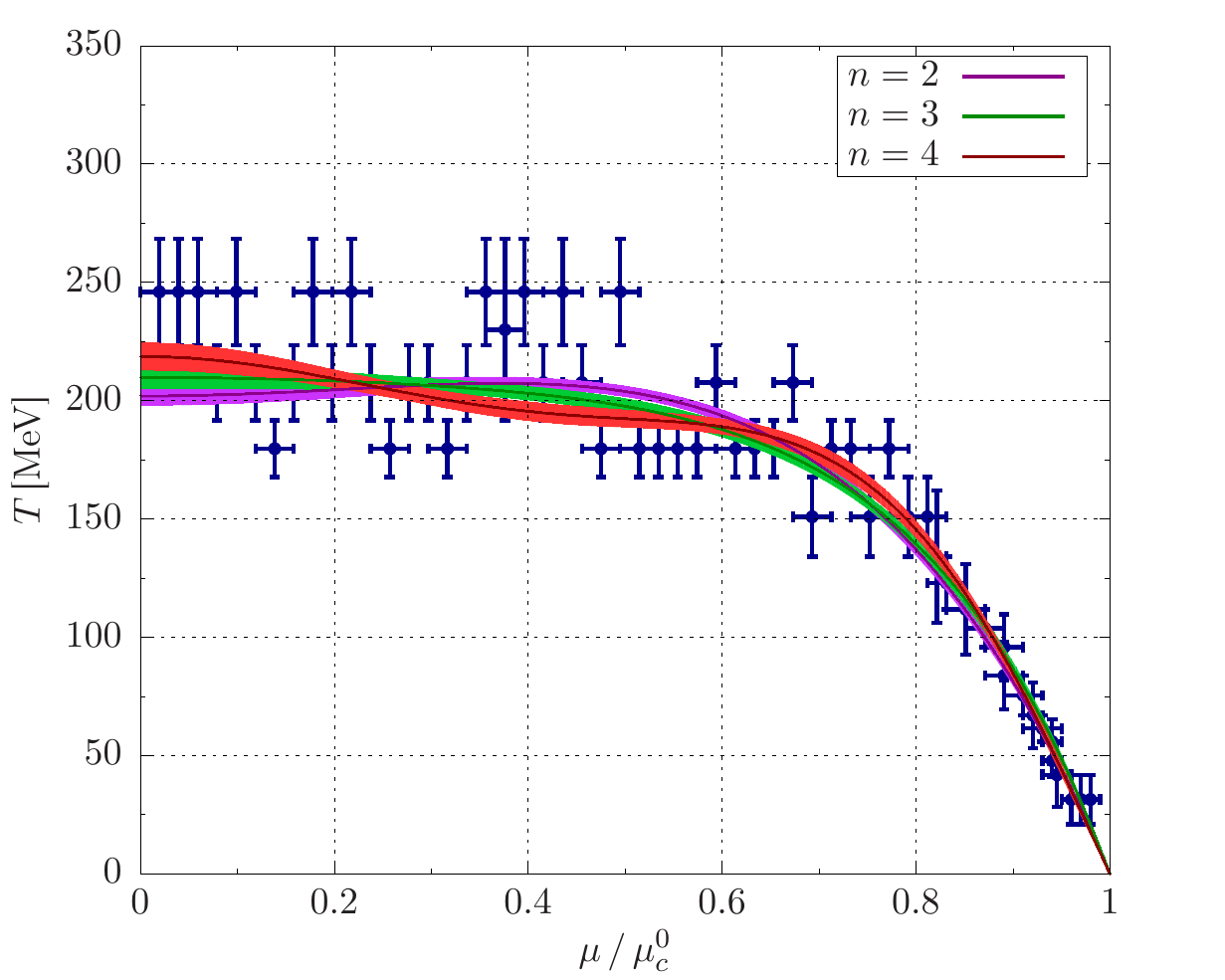}
	\end{tabular}
	\caption{\label{fig.phase.Boundary} Estimates of the phase boundary for QCD 
	in the presence of heavy quarks on a $10^3$ lattice. 
	Left: Comparison of three different fit functions, A, B and C, using $n=2$. 
	Right: Comparison of different orders for fit B. 
	}
\end{figure}

The Binder cumulant for the symmetrised Polyakov loop expectation value $\bra P^{\rm s}\ket$ is shown in Figure~\ref{fig.Binder}. The separation between the confined phase, with $\bra P^{\rm s}\ket=0$, and deconfined phase, with $\bra P^{\rm s}\ket\neq 0$, is clearly visible. At low temperature, the transition can easily be identified, due to the adequate coverage of the parameter space and the relatively sharp transition. At higher temperature, the setup with fixed lattice spacing does not have sufficient resolution to determine the thermal transition with precision. Nevertheless, a clear phase boundary is seen to emerge. 
 To identify the transition between both phases, we determine the parameters for which the Binder cumulant reaches $1/3$, and the results are shown in Fig.~\ref{fig.phase.Boundary}. The uncertainties are estimated by taking half
 the distance between neighbouring points in both $T$ and $\mu$ directions. As mentioned above, the resolution in the temperature direction is limited due to having only integer $N_\tau$ values, which leads to large discretisation effects for the thermal transition.
 The transition to higher densities can be mapped out with much more precision. 

To parameterise the transition temperature as a function of the chemical potential, we have fitted the estimates for $T_c(\mu)$ to a number of fitting functions. Using the notation
\be
x = \left(\frac{\mu}{\mu_c^0}\right)^2,
\ee
we considered an expansion around $x\sim 0$, i.e.,
\begin{equation}
	\label{eq.fit1}
	\mbox{fit A:} \qquad T_c(\mu) = \sum_{k=0}^n a_k x^k,
\end{equation} 
where we used that $T_c(\mu)$ is an even function of $\mu$ \cite{deForcrand:2002hgr}.
Given that due to the lattice setup the transition is better determined around $x\lesssim 1$ than around 0, and that $T_c(\mu_c^0)=0$, we have considered a power series around $x=1$ as well, namely
\begin{equation}
	\label{eq.fit2}
	\mbox{fit B:} \qquad T_c(\mu) = \sum_{k=1}^n b_k (1-x)^k.
\end{equation} 
The expansion parameters $\{a_k\}$ and $\{b_k\}$ are trivially related, provided that $\sum_k a_k=0$ emerges from the fit.
Finally, to take into account nonanalytic behaviour around $x=1$, as required by the Clausius-Clapeyron relation ($\partial T_c(\mu)/\partial \mu\to \infty$ at $\mu=\mu_c^0$), we included one additional term and used
\begin{equation}
	\label{eq.fit3}
	\mbox{fit C:} \qquad T_c(\mu) = c_0(1-x)^\alpha + \sum_{k=1}^n c_k (1-x)^k,
\end{equation} 
with $0<\alpha<1$.

\begin{table}[t] 

\begin{tabular}{c||c|c|c|c||c|c|c}
	& \multicolumn{4}{c||}{fit A, $n=2$} & \multicolumn{3}{c}{fit B, $n=2$} \\
	$V$ & $a_0$ & $a_1$ & $a_2$ & $\chi^2_{\text{red}}$ & $b_1$ & $b_2$ &$\chi^2_{\text{red}}$\\
	\hline
	$6^3$ & 276.9 (7.2) & 7.4 (33.7) & -283.4 (31.8)
	& 0.85 & 564.3 (15.2) & -287.8 (19.2) & 0.83\\
	$8^3$ & 216.4 (5.0) & 86.0(25.5) & -305.8 (24.8)
	& 1.51 & 507.8 (12.8) & -289.9 (15.7) & 1.49\\
	$10^3$ & 203.9 (4.3) & 58.9 (23.1) & -257.1 (23.2) & 1.62 &
	481.4 (12.4) & -279.3 (15.0) & 1.62
\end{tabular}

\caption{\label{tb.BoundaryFits} Fit parameters and reduced $\chi^2$ for fits A
and B, see Eqs.\ (\ref{eq.fit1}, \ref{eq.fit2}), used to describe the chemical potential dependence of the transition temperature, $T_c(\mu)$, for three spatial volumes.
}
\end{table}

\begin{figure}[!ht] 
 \begin{center}
 \includegraphics[scale=0.60,clip, trim=0.1cm 0cm 0cm
	0cm]{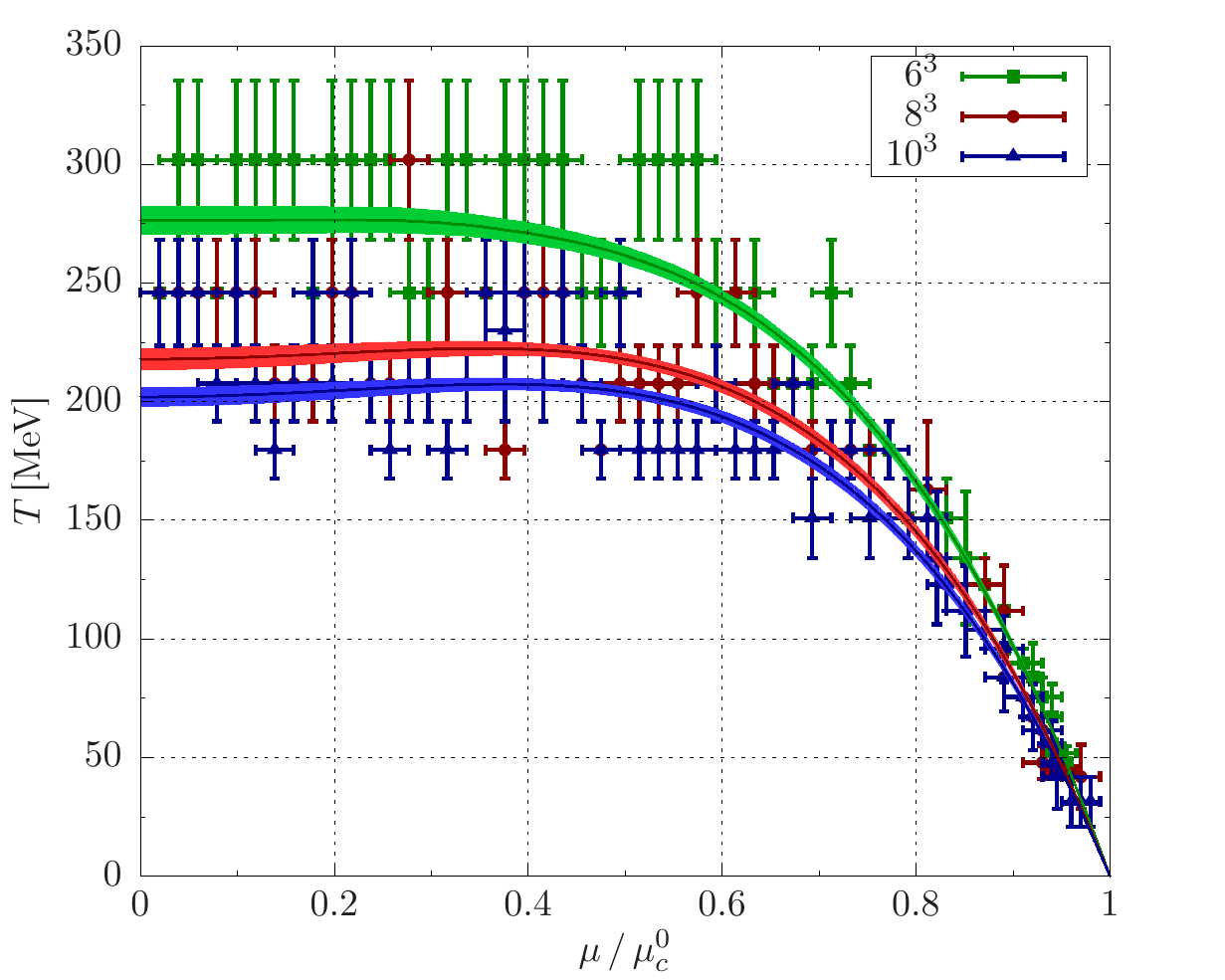} 
	 \end{center}
	\caption{\label{2fig.phase.Boundary} Volume dependence of the phase boundary,
	using fit B with $n=2$.
	}
\end{figure}

The left panel of Fig.~\ref{fig.phase.Boundary} shows fits A, B and C with $n=2$ for our largest volume of $V=10^3$.
The non-analytic behaviour at $\mu \approx \mu^0_c$ is not evident from our data; our lowest temperature is still away from 0. Hence treating $\alpha$ in fit C as a fit parameter does not yield additional information on the transition line and we do not consider C any further.
Fits A and B are seen to be compatible with each other, indicating that $T_c(\mu^0_c) = 0$ emerges without imposing it.
The fit coefficients and the corresponding reduced $\chi^2$ can be found in Table~\ref{tb.BoundaryFits} for fit A and B, for $n=2$.
Note that a rough estimate for $T_c(\mu=0)$ in MeV is given by $a_0\sim b_1+b_2$, which sets the scale of the coefficients.

On the right-hand side of Fig.~\ref{fig.phase.Boundary} we compare three different polynomials for fit B with $n=2, 3$ and $4$, again for $V=10^3$. Higher-orders polynomials result in an almost identical curve as the fourth-order polynomial fit ($n=2$). Hence adding more parameters does not result in an improved fit.
Fits B with $n=2$ for all three volumes studied here are shown in Fig.~\ref{2fig.phase.Boundary}. We observe clear finite-size effects, especially for the smallest simulation box ($6^3$). A much smaller trend can be seen in the two larger volumes. The main limitation, however, comes from the discretisation at high temperature, as discussed above.

The Binder cumulant is in principle suitable to determine the order
of the phase transition, as its value at the transition point only depends on
the universality class~\cite{Binder:1981sa}. Further analyses of the volume
dependence would, however, require a more precise determination of $T_c$
as a function of $\mu$ throughout the phase diagram, with smaller uncertainties.

\section{\label{sec.insta}Instabilities}

In our simulations we encounter instabilities, complicating the analysis. 
These result in a widening of the distribution of observables during the Langevin process
 and affect susceptibilities and other quantities significantly. Based on the formal justification \cite{Aarts:2009uq,Aarts:2011ax} and a comparison with reweighting \cite{DePietri:2007ak}, one can conclude that the wider distributions do not reflect the original theory. In this section we describe some of these features.

\begin{figure}[!h]
	\begin{tabular}{cc}
	\includegraphics[scale=0.57]{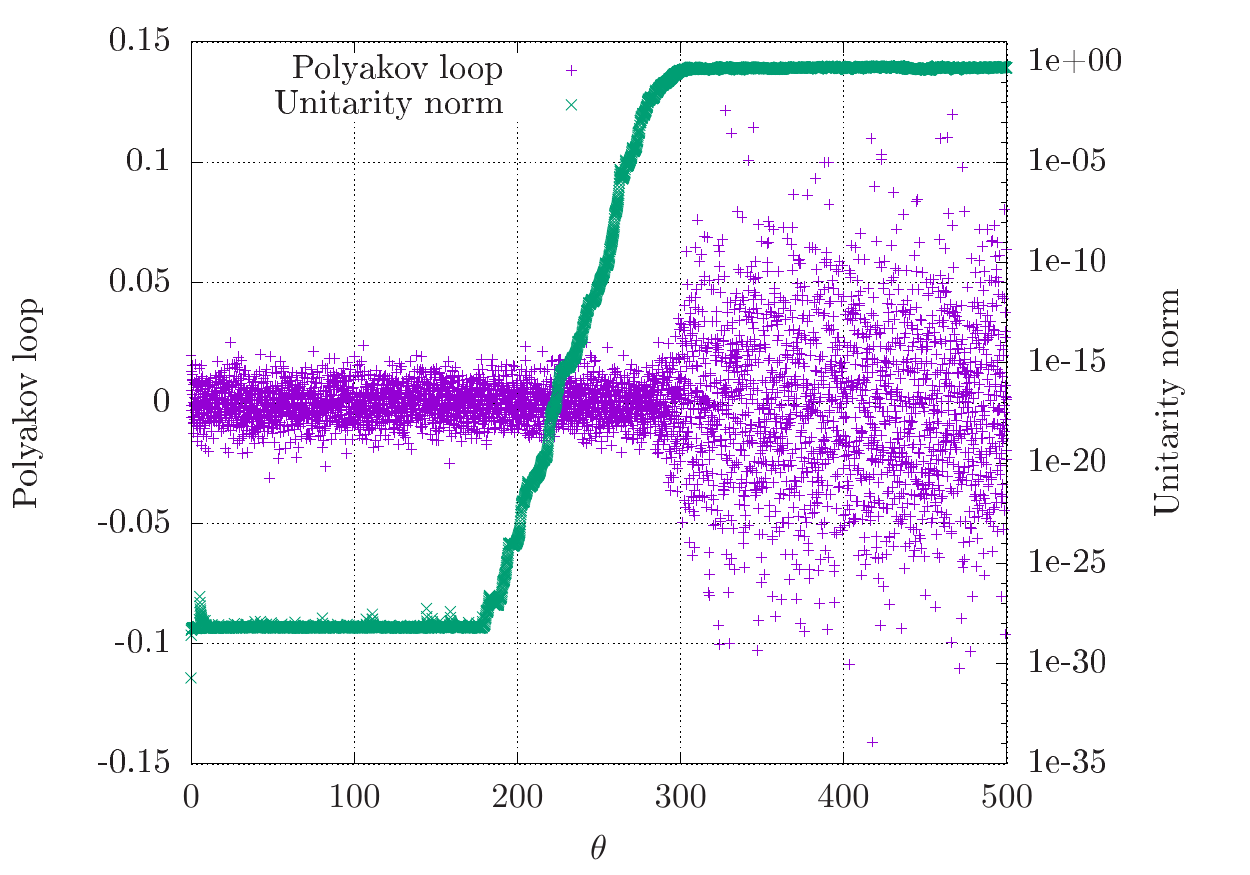}&
	\includegraphics[scale=0.57]{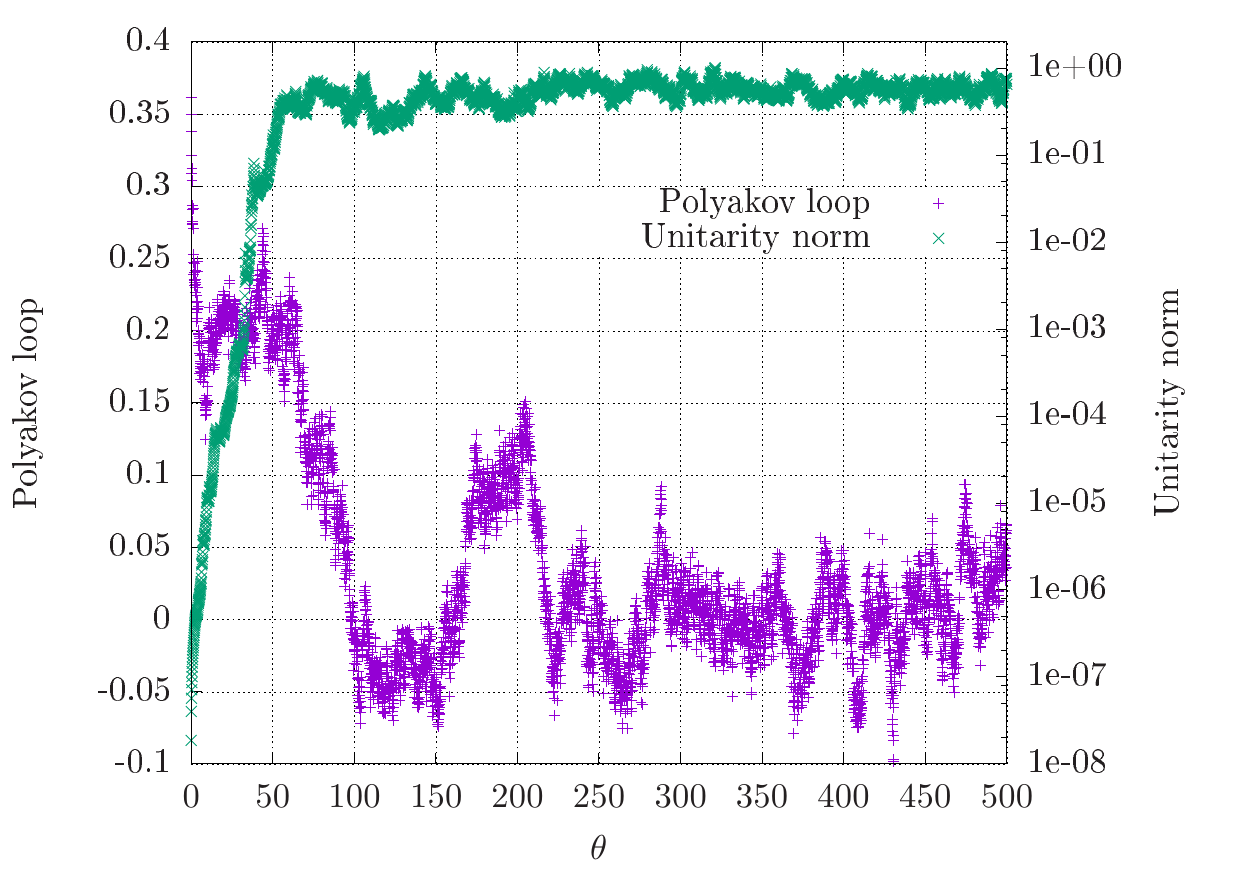}
	\end{tabular}
	\caption{\label{fig.widen.1} Real part of the Polyakov loop $P$ and unitarity norm $d_2$ as a function of Langevin time
	$\theta$ at low temperature ($N_\tau=20, \mu=0.5$, left) and high temperature ($N_\tau=4, \mu=0.7$, right) on a $10^3$ lattice.
		}
\end{figure}

In Fig.\ \ref{fig.widen.1} we show examples of the Langevin time evolution of the real part of the Polyakov loop $P$ and the unitarity norm $d_2$, at low (left) and high (right) temperature. We observe two distinct segments, characterised by a small unitarity norm and controlled fluctuations in the initial part, followed by larger fluctuations and unitarity norm afterwards. At the higher temperature, this also leads to a tunnelling transition for the Polyakov loop, from around 0.2 to 0. 

\begin{table}[!htb]
\begin{center}
\begin{tabular}{|c|c|c|c|}
	\hline
	& \multicolumn{3}{c|}{$N_\tau = 20,\,\, \mu =0.5$} \\
	\hline
	& $100 < \theta < 250$ & $330 < \theta < 500$ & Reweighting \\ 
	\hline
	$\bra P\ket$ & $0.00009(65)$ & $-0.0002(44)$ & $0.000032(22)$ \\
	$\chi_P$ & $0.0542(68)$ & $0.0510(1796)$ & $0.055473(68)$ \\
	$B$ & $0.01(17)$ & $-22(207)$ & $0.0013(19)$ \\
	\hline
	\hline
	& \multicolumn{3}{c|}{$N_\tau = 4,\,\, \mu =0.7$} \\
	\hline
	& $20 < \theta < 60$ & $100 < \theta < 500$ & Reweighting \\
	\hline
	$\bra P\ket$  & $0.2043(53)$ &
	$0.0069(115)$ & $0.202717(66)$ \\
	$\chi_P$ & $0.37(17)$ &
	$1.44(73)$ & $0.37993(17)$\\
	$B$ & $0.6544(57)$ &
	$-0.6332(8105)$ & $0.65487(18)$ \\
	\hline
\end{tabular}
\end{center}
\caption{\label{tb.HD.10.20.Stat} 
	Analysis of the real part of the Polyakov loop, its susceptibility and Binder cumulant, for the data presented in Fig.~\ref{fig.widen.1}.
	In each case, the two intervals correspond to the regions where the
	Polyakov loop fluctuations are consistent around a given value.
	Reweighting results are added for comparison.
	} 
\end{table}

\begin{figure}[!h]
	\begin{tabular}{cc}
	\includegraphics[scale=0.57]{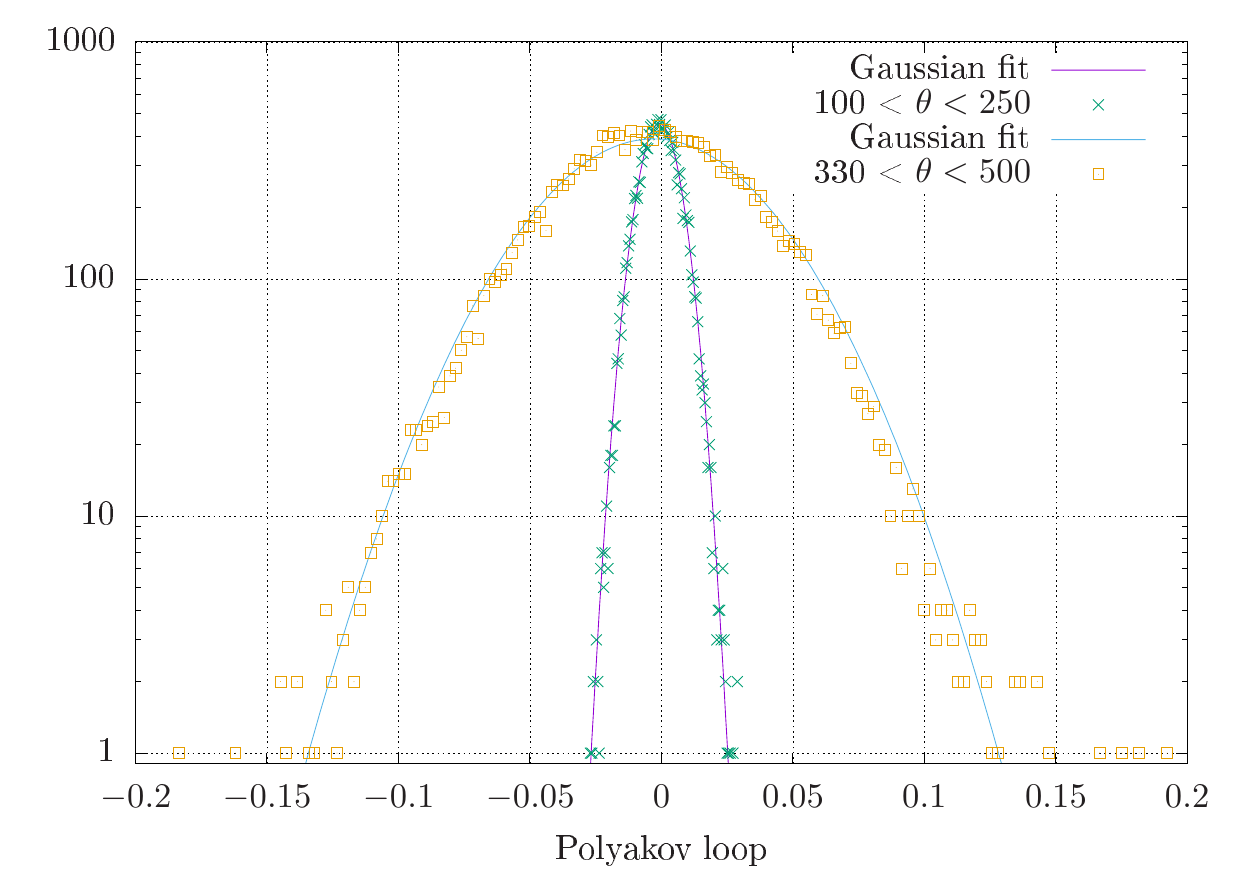}&
	\includegraphics[scale=0.57]{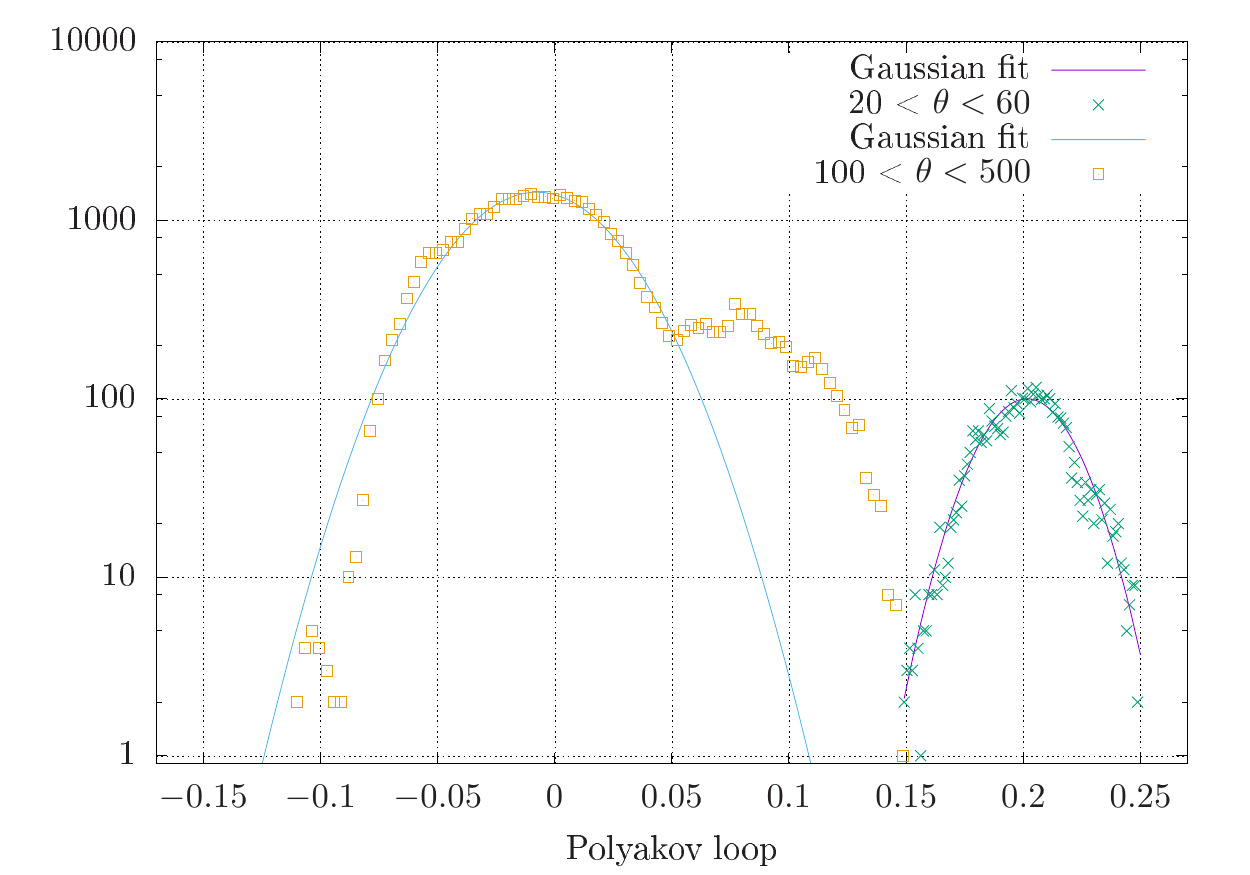}
	\end{tabular}
	\caption{\label{fig.tunnel.hist} Histograms of the real part of the Polyakov loop before and
	after the rise of the unitarity norm, for the runs presented in Fig.~\ref{fig.widen.1},
	at low (left) and high (right) temperature.}
\end{figure}

The data in Fig.~\ref{fig.widen.1} is analysed further in Table~\ref{tb.HD.10.20.Stat}.
We have determined expectation values for the Polyakov loop, and its susceptibility $\chi_P$ and Binder cumulant in each of the two intervals where the Polyakov loop fluctuates consistently around a certain value.
Results obtained with reweighting are shown as well. We note that the observables are, within the statistical error, in agreement with the latter in the first interval, but not in the second one.
The apparent agreement $B\sim 0$ at low temperature for the entire interval mostly reflects that $\bra P\ket\sim 0$ throughout, and hence the susceptibility is a more sensitive measure of accuracy. 
In Fig.~\ref{fig.tunnel.hist} we compare histograms for both
scenarios. A Gaussian fit is added to guide the eye. For the region 
with larger unitarity norms, the distribution is broader, with a larger tail.
At high temperature, there is in addition a shift of the mean.
We conclude that the region with smaller unitarity norm leads 
to acceptable results, while those with a larger value do not.

The behaviour described above has been seen for different chemical potentials and
temperatures, but in all cases widening of the distributions coincided with a severe 
change in unitarity norm. We have checked that using smaller stepsizes does not
prevent these transition from occurring.
 The inability to control the unitarity norm on coarser lattices was
already noted in Ref.\ \cite{Seiler:2012wz}.

\begin{figure}[!h]
	\begin{tabular}{cc}
	\includegraphics[scale=0.57]{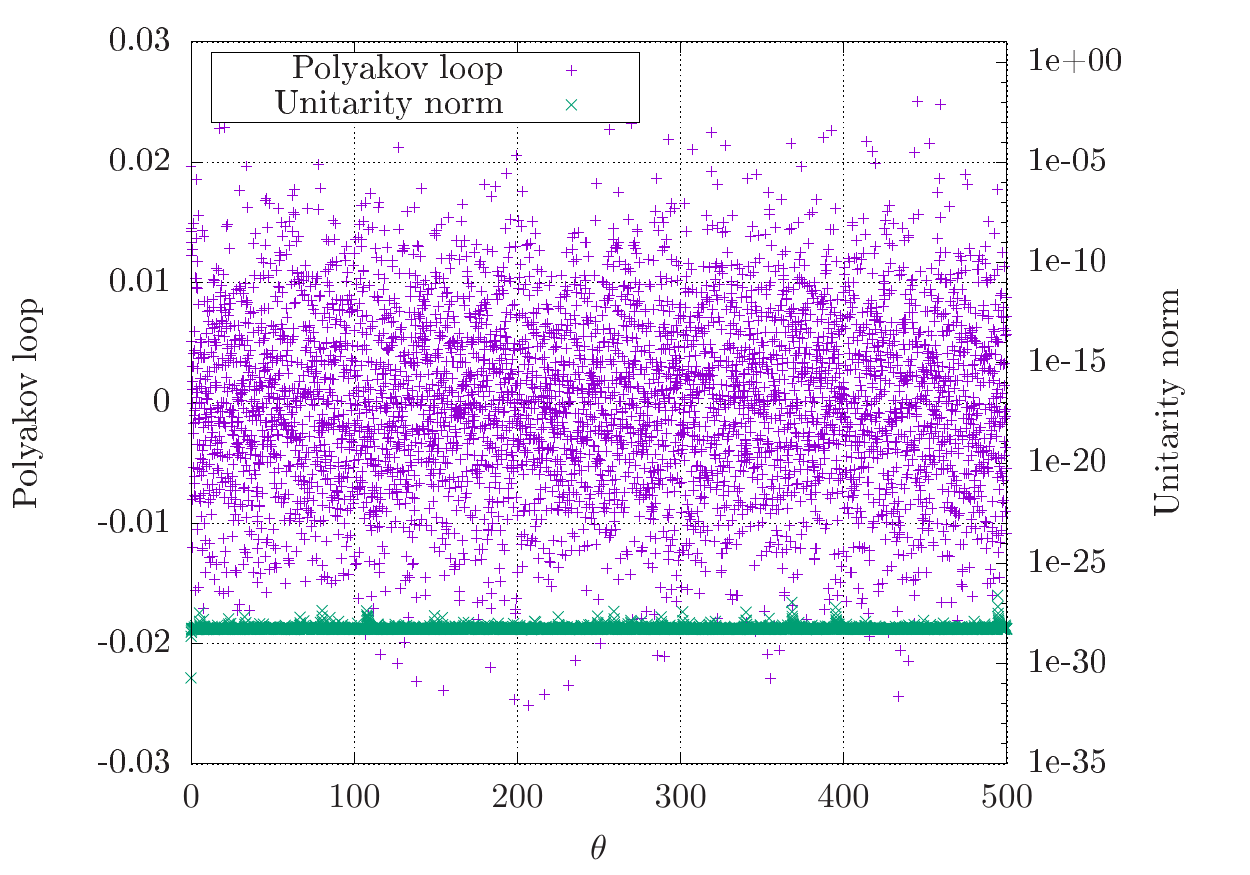}&
	\includegraphics[scale=0.57]{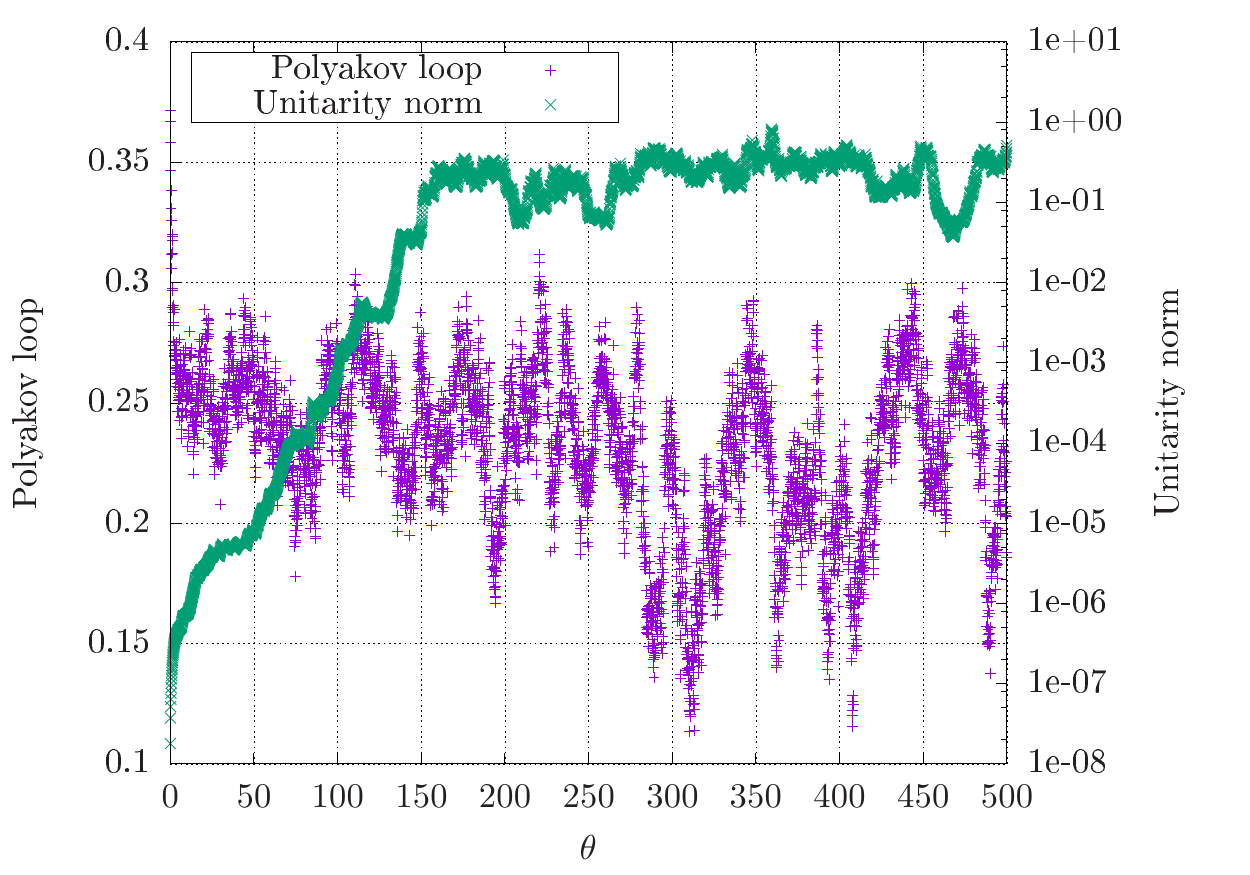}
	\end{tabular}
	\begin{tabular}{cc}
	\includegraphics[scale=0.57]{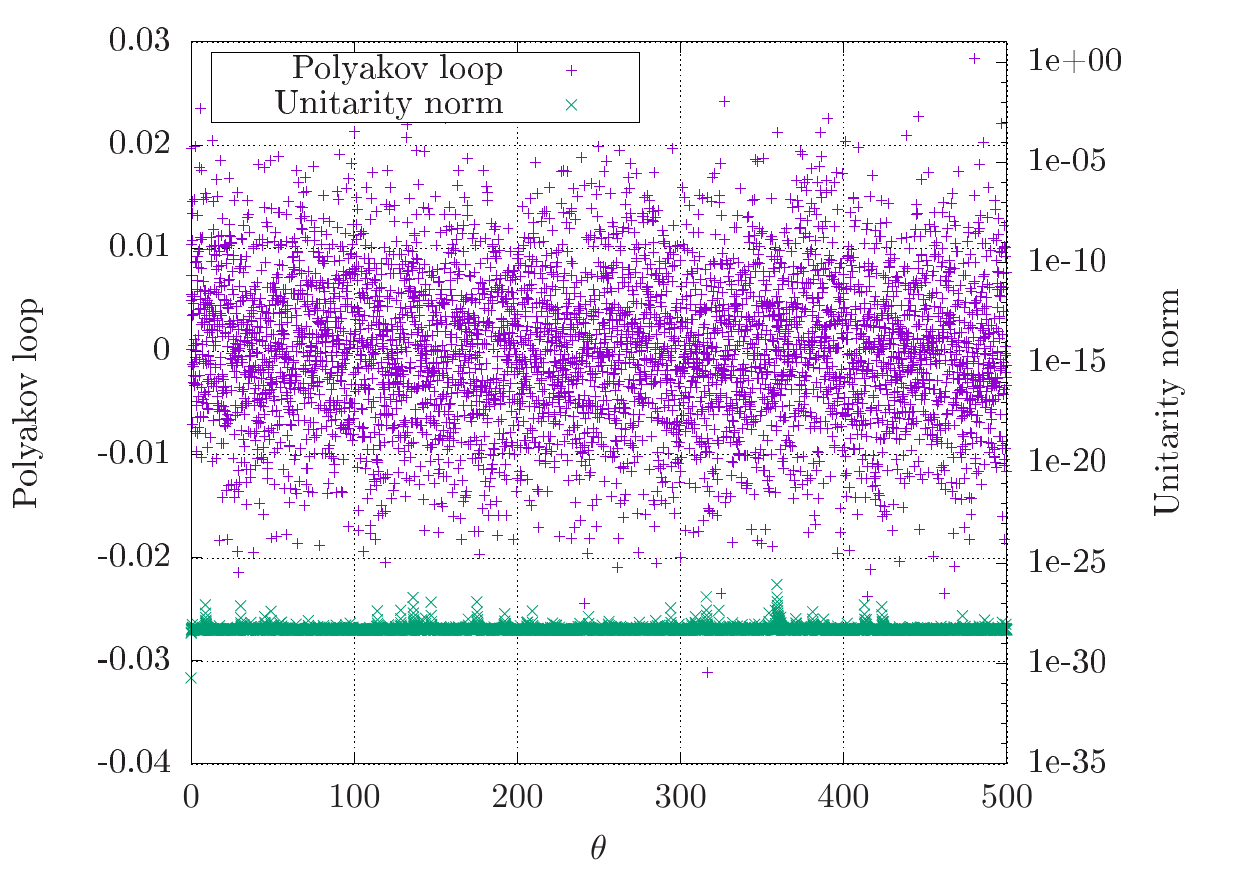}&
	\includegraphics[scale=0.57]{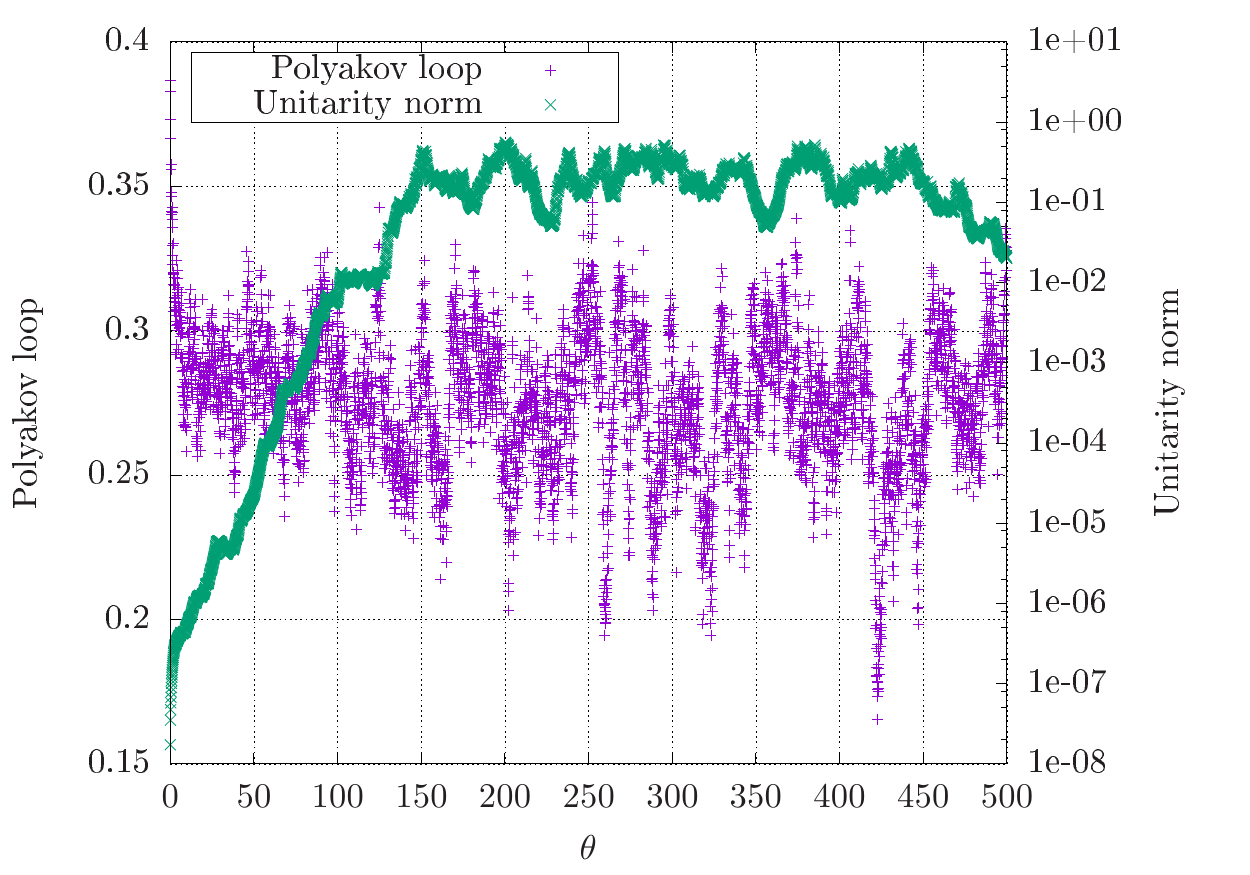}
	\end{tabular}
	\caption{\label{fig.instabilities.beta} Real part of the Polyakov loop $P$ and
	unitarity norm $d_2$ for a larger gauge coupling of $\beta = 6.0$ (top) and
	$\beta = 6.2$ (bottom) with low temperature ($N_\tau=20, \mu=0.5$, left) and
	high temperature ($N_\tau=4, \mu=0.7$, right) on a $10^3$ lattice. }
\end{figure}

\begin{figure}[!h]
	\begin{tabular}{cc}
	\includegraphics[scale=0.57]{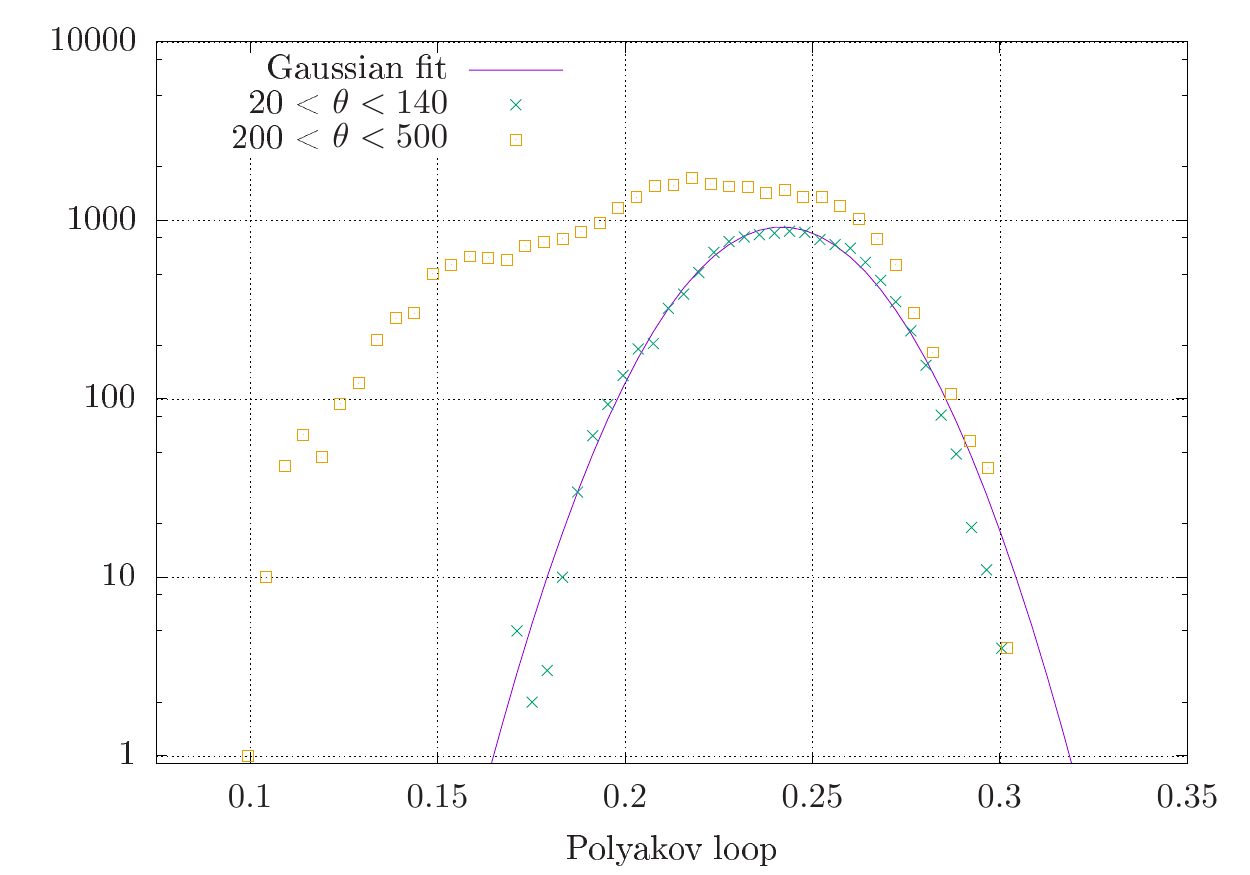}&
	\includegraphics[scale=0.57]{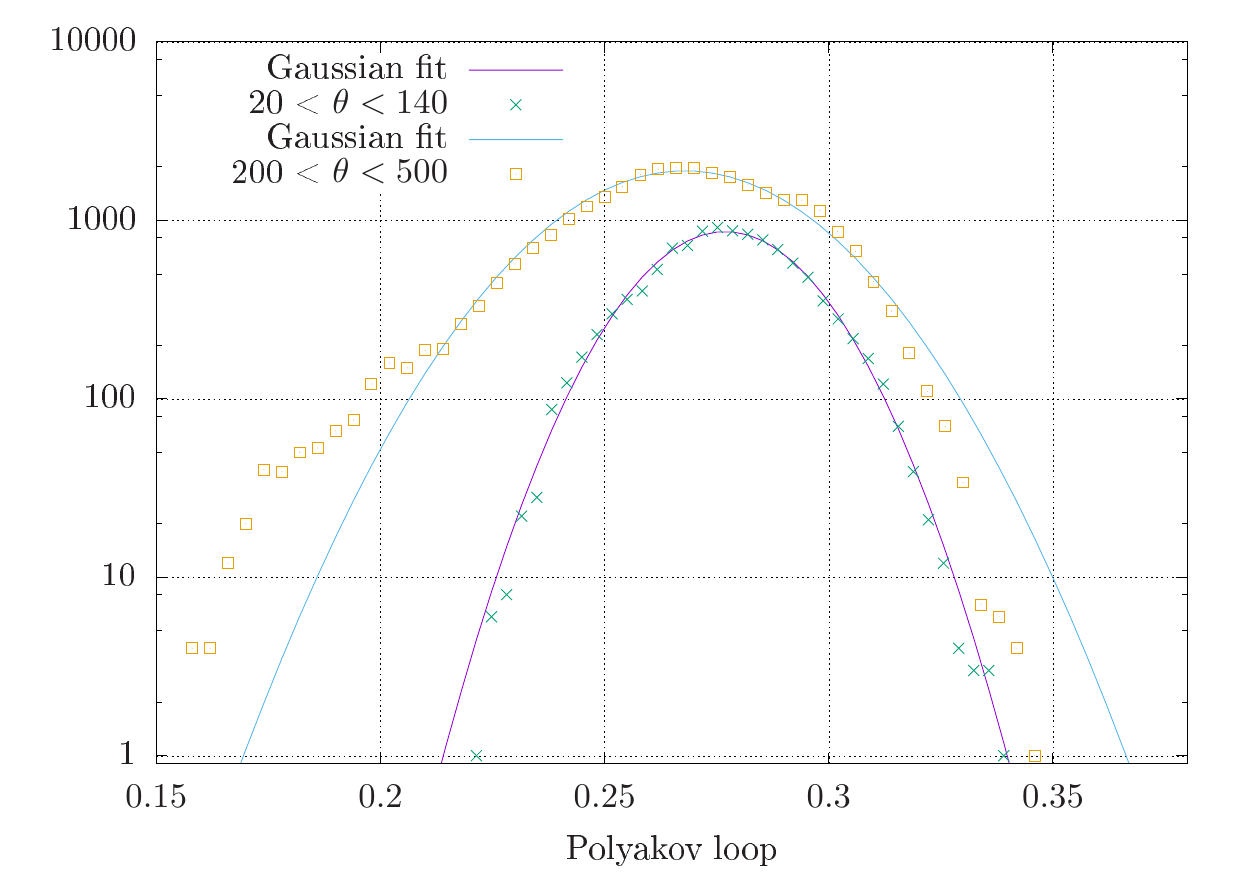}
	\end{tabular}
	\caption{\label{fig.histo.beta} Histograms of the real part of the
	Polyakov loop before and after the rise of the unitarity norm, for the larger gauge couplings
	of $\beta = 6.0$ (left) and $6.2$ (right), at high temperature 
	($N_\tau=4, \mu=0.7$).
	}
\end{figure}

To check the behaviour closer to the continuum limit, we have performed additional 
simulations with larger gauge coupling, $\beta = 6.0$ and $6.2$. Fig.~\ref{fig.instabilities.beta}
shows the real part of the Polyakov loop for an identical setup as in
Fig.~\ref{fig.widen.1} and Table~\ref{tb.HD.10.20.Stat}.
Simulations at low temperature ($N_\tau=20$) are shown on the left and at high temperature ($N_\tau=4$) on the right. 
On the finer lattices and at low temperature, the unitarity norms remain practically $0$ for the entire simulation time.
At the higher temperature, the unitarity norm still rises, but with a smaller exponent.
Once the unitarity norm becomes too large, fluctuations become significantly larger and skirts emerge, as in the case discussed above.
This behaviour can be seen in Fig.~\ref{fig.histo.beta}, which shows the histograms for the high-temperature runs for the two larger $\beta$ values on a $10^3$ lattices. 
Hence we conclude that the instabilities are still present on finer lattices, but that they set in later (at high temperature) or only appear beyond the length of the Langevin trajectory (at low temperature).

In order to maintain the volume of the lattice in physical units, simulations with a larger gauge coupling require larger simulation volumes to compensate the smaller lattice spacing. In the preceding section, we have found that employing a gauge coupling of $\beta = 5.8$ yields a compromise between simulation costs and the ability to extract reliable information on the phase boundary of HDQCD.
 The Langevin time when the unitarity norm starts rising varies considerably
for different setups. In most cases that happens sufficiently 
after the thermalisation stage, which leaves enough data points to
allow us to extract observables and perform a subsequent analysis. 
However, since the amount of available
data suitable for analysis differs greatly between ensembles, we find
different uncertainties in each setup, including the integrated auto-correlation
time~\cite{Wolff:2003sm}.
In order to implement these findings, we have made sure that for the 
results presented in the previous section, only simulation data for 
which the unitarity norm $d_2$ is smaller than 0.03 were included.

\begin{figure}[!h]
	\begin{tabular}{cc}
	\includegraphics[scale=0.57]{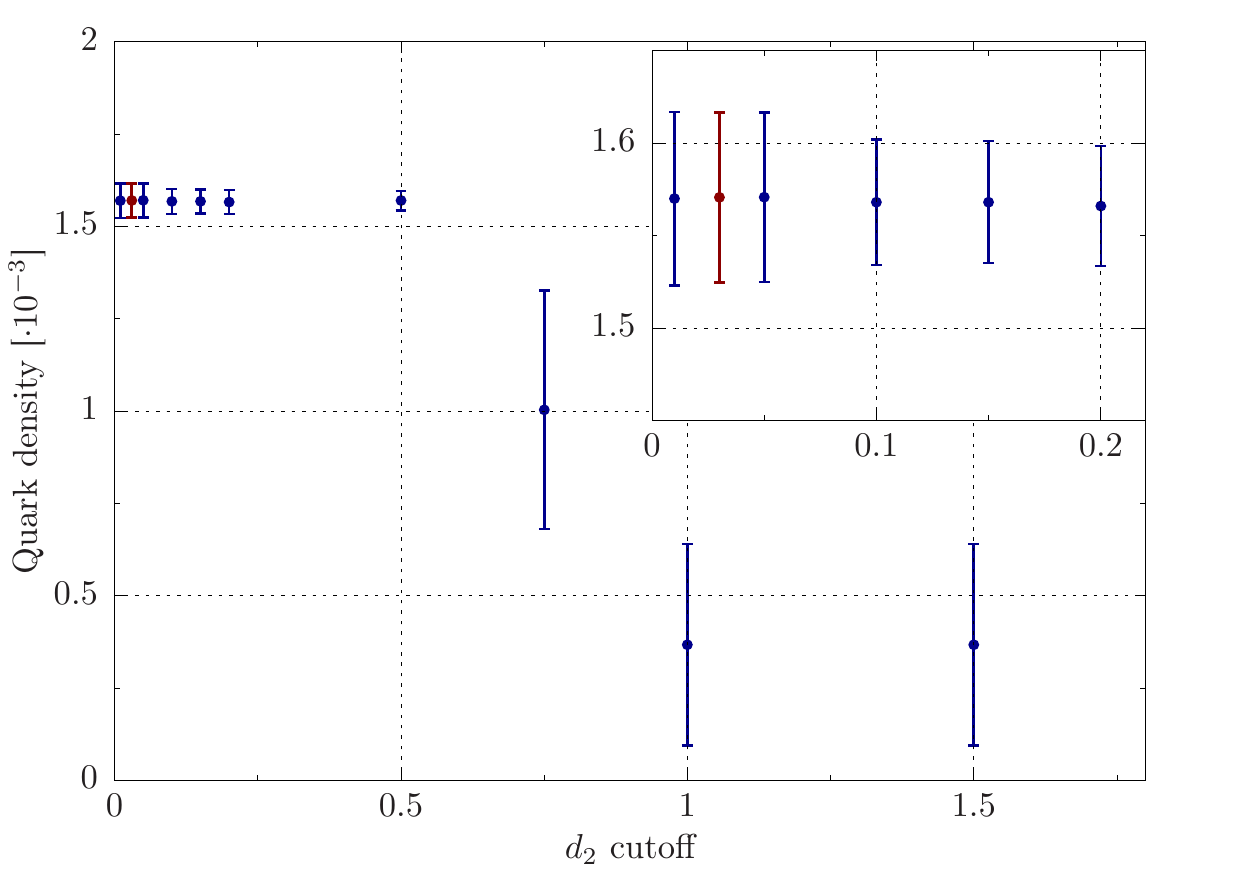}&
	\includegraphics[scale=0.57]{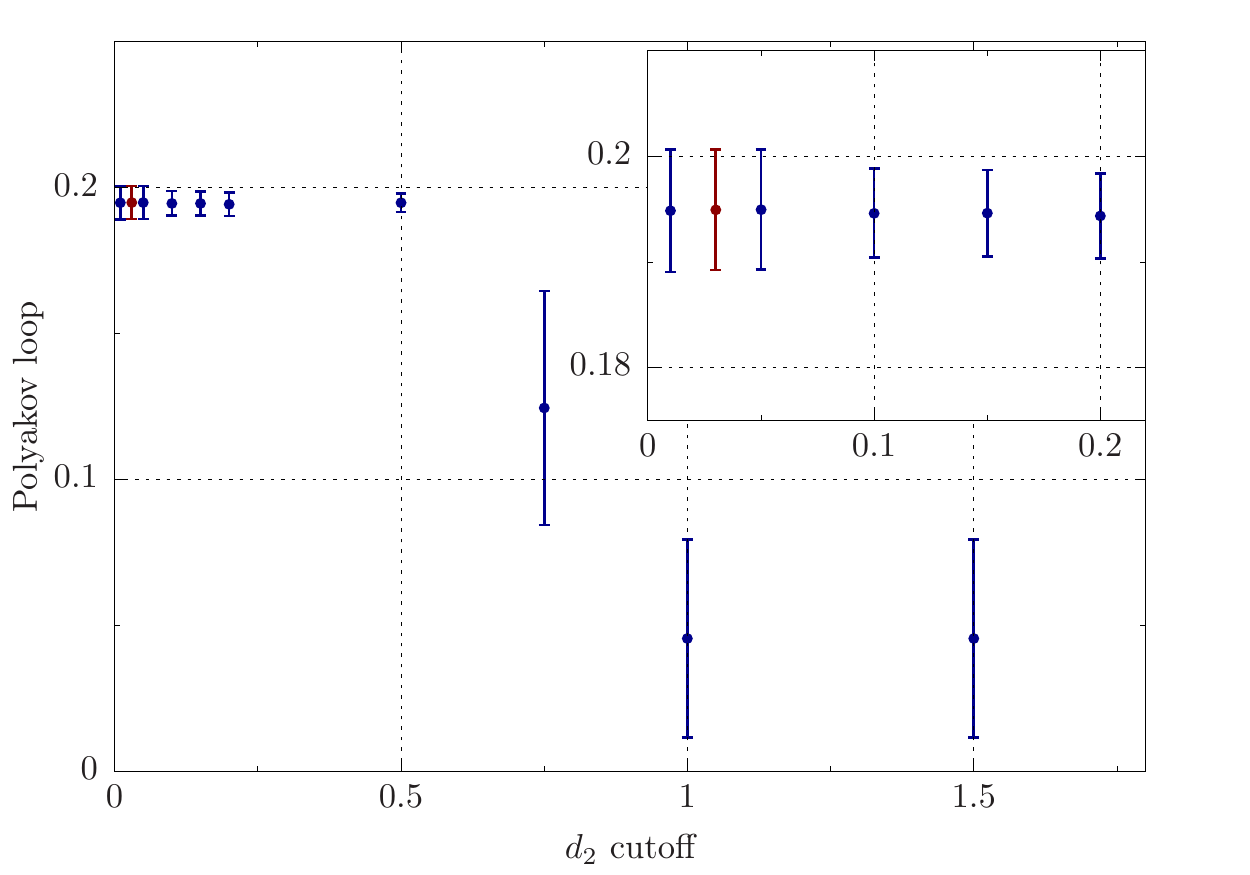}
	\end{tabular}
	\caption{\label{fig.instabilities.cutoff} The quark density (left) and the
	Polyakov loop (right) as a function of the cutoff imposed on the unitarity norm
	$d_2$ for $N_\tau = 4,\mu =0.7$ on a $10^3$ lattice. For small
	unitarity norms, $d_2<0.5$, the observables are independent of the cutoff. The
	red point indicates the value chosen in this study. The insets focus on the
	region of smaller cutoffs.}
\end{figure}

To check the sensitivity with respect to changes in the cutoff imposed on
the unitarity norm $d_2$ and the robustness of physical observables, we
show in Fig.\ 10 the dependence of the quark density and the Polyakov loop
on the maximally allowed unitarity norm, for $N_\tau=4$ and $\mu=0.7$ on a
$10^3$ lattice. We observe that the obervables are stable and independent
of the cutoff over a wide range, up to $d_2\sim 0.5$. A clear transition
is visible for larger values of the cutoff, which coincides with the
widening of the distributions, discussed above. Note that the change in
the statistical uncertainties at larger $d_2$ cutoff follows from the
decrease of data points available in the analysis. Similar behaviour is seen at
other parameter values. We conclude that observables are robust under changes in
the cutoff imposed on the unitarity norm. In the previous section we have
conservatively chosen a small value for the cutoff, i.e.\ $d_2<0.03$, to
stay sufficiently away from the wrong behaviour observed for unitarity
norm of ${\cal O}(1)$.

\section{\label{sec.concl}Conclusion}

We have studied the phase diagram of QCD in the presence of heavy quarks, using complex Langevin simulations. Combining gauge cooling with a careful monitoring of the Langevin process, we have shown that it is possible to perform ab-initio simulations in the entire $T-\mu$ plane. The phase boundary between the confined and the deconfined phases was determined via the Binder cumulant of the symmetrised Polyakov loop and the resulting line can be fitted in terms of simple polynomials. In our setup, in which the lattice spacing is fixed and temperature is varied by changing the temporal extent, the main uncertainty occurs at high temperature, where discretisation effects are severe. The transition at low temperature, however, can be determined with more precision.

During the Langevin process, we observed instabilities, which take configurations far away from the SU(3) submanifold, even in the presence of gauge cooling. These events lead to incorrect convergence. By monitoring the unitarity norm, we found that it is nevertheless possible to collect sufficient simulation data which can be used in a reliable manner. There are strong indications that this situation will improve on finer lattices.

As an outlook, we note that in order to determine the phase boundary, and the order of the transition, throughout the $T-\mu$ plane with more precision, it will be necessary to vary both the lattice spacing and the temporal extent simultaneously, both in the model considered here as in full QCD. Besides this, an important additional step is a better control on the Langevin process and work in this direction is currently under development.

\acknowledgments
The authors would like to thank Erhard Seiler and Ion-Olimpiu Stamatescu for valuable discussions and collaboration, and contributions with regard to the reweighting method. 
We are grateful for the computing resources made available by HPC Wales. 
This work used the DiRAC Blue Gene Q Shared Petaflop system at the University of Edinburgh, operated by the Edinburgh
Parallel Computing Centre on behalf of the STFC DiRAC HPC Facility
(www.dirac.ac.uk). This equipment was funded by BIS National E-infrastructure
capital grant ST/K000411/1, STFC capital grant ST/H008845/1, and STFC DiRAC
Operations grants ST/K005804/1 and ST/K005790/1. DiRAC is part of the National
E-Infrastructure. We acknowledge the STFC grant ST/L000369/1, the Royal Society
and the Wolfson Foundation.
We also thank the Gauss Centre for Supercomputing
e.V. (www.gauss-centre.eu) for providing computing
time on the GCS Supercomputer SuperMUC at Leibniz Supercomputing
Centre (LRZ), as well as providing computing
time through the John von Neumann Institute for
Computing (NIC) on the GCS share of the supercomputer JURECA and
JUQUEEN \cite{juqueen} at J\"ulich Supercomputing Centre (JSC).
FA is grateful for the support through the Brazilian
government programme ``Science without Borders'' under scholarship number BEX
9463/13-5.

\bibliography{hdqcd-v4}{}

\providecommand{\href}[2]{#2}\begingroup\raggedright\begin{thebibliography}{10}

\bibitem{Borsanyi:2013bia}
S.~Bors\'anyi, Z.~Fodor, C.~Hoelbling, S.~D. Katz, S.~Krieg and K.~K. Szabo,
  \emph{{Full result for the QCD equation of state with 2+1 flavors}},
  \href{http://dx.doi.org/10.1016/j.physletb.2014.01.007}{\emph{Phys. Lett.}
  {\bf B730} (2014) 99--104}, [\href{http://arxiv.org/abs/1309.5258}{{\tt
  1309.5258}}].

\bibitem{Bazavov:2014pvz}
{\scshape HotQCD} collaboration, A.~Bazavov et~al., \emph{{Equation of state in
  ( 2+1 )-flavor QCD}},
  \href{http://dx.doi.org/10.1103/PhysRevD.90.094503}{\emph{Phys. Rev.} {\bf
  D90} (2014) 094503}, [\href{http://arxiv.org/abs/1407.6387}{{\tt
  1407.6387}}].

\bibitem{deForcrand:2010ys}
P.~de~Forcrand, \emph{{Simulating QCD at finite density}}, {\emph{PoS} {\bf
  LAT2009} (2009) 010}, [\href{http://arxiv.org/abs/1005.0539}{{\tt
  1005.0539}}].

\bibitem{Aarts:2013lcm}
G.~Aarts, \emph{{Complex Langevin dynamics and other approaches at finite
  chemical potential}}, {\emph{PoS} {\bf LATTICE2012} (2012) 017},
  [\href{http://arxiv.org/abs/1302.3028}{{\tt 1302.3028}}].

\bibitem{Sexty:2014dxa}
D.~Sexty, \emph{{New algorithms for finite density QCD}}, {\emph{PoS} {\bf
  LATTICE2014} (2014) 016}, [\href{http://arxiv.org/abs/1410.8813}{{\tt
  1410.8813}}].

\bibitem{Scorzato:2015qts}
L.~Scorzato, \emph{{The Lefschetz thimble and the sign problem}},  in
  \emph{{Proceedings, 33rd International Symposium on Lattice Field Theory
  (Lattice 2015)}}, 2015.
\newblock \href{http://arxiv.org/abs/1512.08039}{{\tt 1512.08039}}.

\bibitem{Gattringer:2016kco}
C.~Gattringer and K.~Langfeld, \emph{{Approaches to the sign problem in lattice
  field theory}}, {\emph{International Journal of Modern Physics A} (2016) },
  [\href{http://arxiv.org/abs/1603.09517}{{\tt 1603.09517}}].

\bibitem{Cohen:2003kd}
T.~D. Cohen, \emph{{Functional integrals for QCD at nonzero chemical potential
  and zero density}},
  \href{http://dx.doi.org/10.1103/PhysRevLett.91.222001}{\emph{Phys. Rev.
  Lett.} {\bf 91} (2003) 222001},
  [\href{http://arxiv.org/abs/hep-ph/0307089}{{\tt hep-ph/0307089}}].

\bibitem{Bender:1992gn}
I.~Bender, T.~Hashimoto, F.~Karsch, V.~Linke, A.~Nakamura, M.~Plewnia et~al.,
  \emph{{Full QCD and QED at finite temperature and chemical potential}},
  \href{http://dx.doi.org/10.1016/0920-5632(92)90265-T}{\emph{Nucl. Phys. Proc.
  Suppl.} {\bf 26} (1992) 323--325}.

\bibitem{deForcrand:2014tha}
P.~de~Forcrand, J.~Langelage, O.~Philipsen and W.~Unger, \emph{{Lattice QCD
  Phase Diagram In and Away from the Strong Coupling Limit}},
  \href{http://dx.doi.org/10.1103/PhysRevLett.113.152002}{\emph{Phys. Rev.
  Lett.} {\bf 113} (2014) 152002}, [\href{http://arxiv.org/abs/1406.4397}{{\tt
  1406.4397}}].

\bibitem{Glesaaen:2015vtp}
J.~Glesaaen, M.~Neuman and O.~Philipsen, \emph{{Equation of state for cold and
  dense heavy QCD}},
  \href{http://dx.doi.org/10.1007/JHEP03(2016)100}{\emph{JHEP} {\bf 03} (2016)
  100}, [\href{http://arxiv.org/abs/1512.05195}{{\tt 1512.05195}}].

\bibitem{DePietri:2007ak}
R.~De~Pietri, A.~Feo, E.~Seiler and I.-O. Stamatescu, \emph{{A Model for QCD at
  high density and large quark mass}},
  \href{http://dx.doi.org/10.1103/PhysRevD.76.114501}{\emph{Phys. Rev.} {\bf
  D76} (2007) 114501}, [\href{http://arxiv.org/abs/0705.3420}{{\tt
  0705.3420}}].

\bibitem{Ejiri:2013lia}
S.~Ejiri, \emph{{Phase structure of hot dense QCD by a histogram method}},
  \href{http://dx.doi.org/10.1140/epja/i2013-13086-7}{\emph{Eur. Phys. J.} {\bf
  A49} (2013) 86}, [\href{http://arxiv.org/abs/1306.0295}{{\tt 1306.0295}}].

\bibitem{Garron:2016noc}
N.~Garron and K.~Langfeld, \emph{{Anatomy of the sign-problem in heavy-dense
  QCD}},  \href{http://arxiv.org/abs/1605.02709}{{\tt 1605.02709}}.

\bibitem{Aarts:2008rr}
G.~Aarts and I.-O. Stamatescu, \emph{{Stochastic quantization at finite
  chemical potential}},
  \href{http://dx.doi.org/10.1088/1126-6708/2008/09/018}{\emph{JHEP} {\bf 09}
  (2008) 018}, [\href{http://arxiv.org/abs/0807.1597}{{\tt 0807.1597}}].

\bibitem{Aarts:2009dg}
G.~Aarts, F.~A. James, E.~Seiler and I.-O. Stamatescu, \emph{{Adaptive stepsize
  and instabilities in complex Langevin dynamics}},
  \href{http://dx.doi.org/10.1016/j.physletb.2010.03.012}{\emph{Phys. Lett.}
  {\bf B687} (2010) 154--159}, [\href{http://arxiv.org/abs/0912.0617}{{\tt
  0912.0617}}].

\bibitem{Seiler:2012wz}
E.~Seiler, D.~Sexty and I.-O. Stamatescu, \emph{{Gauge cooling in complex
  Langevin for QCD with heavy quarks}},
  \href{http://dx.doi.org/10.1016/j.physletb.2013.04.062}{\emph{Phys. Lett.}
  {\bf B723} (2013) 213--216}, [\href{http://arxiv.org/abs/1211.3709}{{\tt
  1211.3709}}].

\bibitem{Aarts:2014bwa}
G.~Aarts, E.~Seiler, D.~Sexty and I.-O. Stamatescu, \emph{{Simulating QCD at
  nonzero baryon density to all orders in the hopping parameter expansion}},
  \href{http://dx.doi.org/10.1103/PhysRevD.90.114505}{\emph{Phys. Rev.} {\bf
  D90} (2014) 114505}, [\href{http://arxiv.org/abs/1408.3770}{{\tt
  1408.3770}}].

\bibitem{Rindlisbacher:2015pea}
T.~Rindlisbacher and P.~de~Forcrand, \emph{{Two-flavor lattice QCD with a
  finite density of heavy quarks: heavy-dense limit and “particle-hole”
  symmetry}}, \href{http://dx.doi.org/10.1007/JHEP02(2016)051}{\emph{JHEP} {\bf
  02} (2016) 051}, [\href{http://arxiv.org/abs/1509.00087}{{\tt 1509.00087}}].

\bibitem{Seiler:2015uwe}
E.~Seiler and I.-O. Stamatescu, \emph{{A note on the Loop Formula for the
  fermionic determinant}},  \href{http://arxiv.org/abs/1512.07480}{{\tt
  1512.07480}}.

\bibitem{Parisi:1984cs}
G.~Parisi, \emph{{On Complex Probabilities}},
  \href{http://dx.doi.org/10.1016/0370-2693(83)90525-7}{\emph{Phys. Lett.} {\bf
  B131} (1983) 393--395}.

\bibitem{Klauder:1983nn}
J.~R. Klauder, \emph{{Stochastic Quantization}},
  \href{http://dx.doi.org/10.1007/978-3-7091-7651-1_8}{\emph{Acta Phys.
  Austriaca Suppl.} {\bf 25} (1983) 251--281}.

\bibitem{Klauder:1983zm}
J.~R. Klauder, \emph{{A Langevin Approach to Fermion and Quantum Spin
  Correlation Functions}},
  \href{http://dx.doi.org/10.1088/0305-4470/16/10/001}{\emph{J. Phys.} {\bf
  A16} (1983) L317}.

\bibitem{Klauder:1983sp}
J.~R. Klauder, \emph{{Coherent State Langevin Equations for Canonical Quantum
  Systems With Applications to the Quantized Hall Effect}},
  \href{http://dx.doi.org/10.1103/PhysRevA.29.2036}{\emph{Phys. Rev.} {\bf A29}
  (1984) 2036--2047}.

\bibitem{Aarts:2008wh}
G.~Aarts, \emph{{Can stochastic quantization evade the sign problem? The
  relativistic Bose gas at finite chemical potential}},
  \href{http://dx.doi.org/10.1103/PhysRevLett.102.131601}{\emph{Phys. Rev.
  Lett.} {\bf 102} (2009) 131601}, [\href{http://arxiv.org/abs/0810.2089}{{\tt
  0810.2089}}].

\bibitem{Aarts:2010gr}
G.~Aarts and K.~Splittorff, \emph{{Degenerate distributions in complex Langevin
  dynamics: one-dimensional QCD at finite chemical potential}},
  \href{http://dx.doi.org/10.1007/JHEP08(2010)017}{\emph{JHEP} {\bf 08} (2010)
  017}, [\href{http://arxiv.org/abs/1006.0332}{{\tt 1006.0332}}].

\bibitem{Aarts:2011zn}
G.~Aarts and F.~A. James, \emph{{Complex Langevin dynamics in the SU(3) spin
  model at nonzero chemical potential revisited}},
  \href{http://dx.doi.org/10.1007/JHEP01(2012)118}{\emph{JHEP} {\bf 01} (2012)
  118}, [\href{http://arxiv.org/abs/1112.4655}{{\tt 1112.4655}}].

\bibitem{Ambjorn:1985iw}
J.~Ambjorn and S.~K. Yang, \emph{{Numerical Problems in Applying the Langevin
  Equation to Complex Effective Actions}},
  \href{http://dx.doi.org/10.1016/0370-2693(85)90708-7}{\emph{Phys. Lett.} {\bf
  B165} (1985) 140}.

\bibitem{Ambjorn:1986fz}
J.~Ambjorn, M.~Flensburg and C.~Peterson, \emph{{The Complex Langevin Equation
  and Monte Carlo Simulations of Actions With Static Charges}},
  \href{http://dx.doi.org/10.1016/0550-3213(86)90605-X}{\emph{Nucl. Phys.} {\bf
  B275} (1986) 375}.

\bibitem{Berges:2006xc}
J.~Berges, S.~Bors\'anyi, D.~Sexty and I.~O. Stamatescu, \emph{{Lattice
  simulations of real-time quantum fields}},
  \href{http://dx.doi.org/10.1103/PhysRevD.75.045007}{\emph{Phys. Rev.} {\bf
  D75} (2007) 045007}, [\href{http://arxiv.org/abs/hep-lat/0609058}{{\tt
  hep-lat/0609058}}].

\bibitem{Berges:2007nr}
J.~Berges and D.~Sexty, \emph{{Real-time gauge theory simulations from
  stochastic quantization with optimized updating}},
  \href{http://dx.doi.org/10.1016/j.nuclphysb.2008.01.018}{\emph{Nucl. Phys.}
  {\bf B799} (2008) 306--329}, [\href{http://arxiv.org/abs/0708.0779}{{\tt
  0708.0779}}].

\bibitem{Aarts:2010aq}
G.~Aarts and F.~A. James, \emph{{On the convergence of complex Langevin
  dynamics: The Three-dimensional XY model at finite chemical potential}},
  \href{http://dx.doi.org/10.1007/JHEP08(2010)020}{\emph{JHEP} {\bf 08} (2010)
  020}, [\href{http://arxiv.org/abs/1005.3468}{{\tt 1005.3468}}].

\bibitem{Pawlowski:2013pje}
J.~M. Pawlowski and C.~Zielinski, \emph{{Thirring model at finite density in
  0+1 dimensions with stochastic quantization: Crosscheck with an exact
  solution}}, \href{http://dx.doi.org/10.1103/PhysRevD.87.094503}{\emph{Phys.
  Rev.} {\bf D87} (2013) 094503}, [\href{http://arxiv.org/abs/1302.1622}{{\tt
  1302.1622}}].

\bibitem{Pawlowski:2013gag}
J.~M. Pawlowski and C.~Zielinski, \emph{{Thirring model at finite density in
  2+1 dimensions with stochastic quantization}},
  \href{http://dx.doi.org/10.1103/PhysRevD.87.094509}{\emph{Phys. Rev.} {\bf
  D87} (2013) 094509}, [\href{http://arxiv.org/abs/1302.2249}{{\tt
  1302.2249}}].

\bibitem{Aarts:2009uq}
G.~Aarts, E.~Seiler and I.-O. Stamatescu, \emph{{The Complex Langevin method:
  When can it be trusted?}},
  \href{http://dx.doi.org/10.1103/PhysRevD.81.054508}{\emph{Phys. Rev.} {\bf
  D81} (2010) 054508}, [\href{http://arxiv.org/abs/0912.3360}{{\tt
  0912.3360}}].

\bibitem{Aarts:2011ax}
G.~Aarts, F.~A. James, E.~Seiler and I.-O. Stamatescu, \emph{{Complex Langevin:
  Etiology and Diagnostics of its Main Problem}},
  \href{http://dx.doi.org/10.1140/epjc/s10052-011-1756-5}{\emph{Eur. Phys. J.}
  {\bf C71} (2011) 1756}, [\href{http://arxiv.org/abs/1101.3270}{{\tt
  1101.3270}}].

\bibitem{Aarts:2013uxa}
G.~Aarts, L.~Bongiovanni, E.~Seiler, D.~Sexty and I.-O. Stamatescu,
  \emph{{Controlling complex Langevin dynamics at finite density}},
  \href{http://dx.doi.org/10.1140/epja/i2013-13089-4}{\emph{Eur. Phys. J.} {\bf
  A49} (2013) 89}, [\href{http://arxiv.org/abs/1303.6425}{{\tt 1303.6425}}].

\bibitem{Mollgaard:2013qra}
A.~Mollgaard and K.~Splittorff, \emph{{Complex Langevin Dynamics for chiral
  Random Matrix Theory}},
  \href{http://dx.doi.org/10.1103/PhysRevD.88.116007}{\emph{Phys. Rev.} {\bf
  D88} (2013) 116007}, [\href{http://arxiv.org/abs/1309.4335}{{\tt
  1309.4335}}].

\bibitem{Splittorff:2014zca}
K.~Splittorff, \emph{{Dirac spectrum in complex Langevin simulations of QCD}},
  \href{http://dx.doi.org/10.1103/PhysRevD.91.034507}{\emph{Phys. Rev.} {\bf
  D91} (2015) 034507}, [\href{http://arxiv.org/abs/1412.0502}{{\tt
  1412.0502}}].

\bibitem{Nishimura:2015pba}
J.~Nishimura and S.~Shimasaki, \emph{{New Insights into the Problem with a
  Singular Drift Term in the Complex Langevin Method}},
  \href{http://dx.doi.org/10.1103/PhysRevD.92.011501}{\emph{Phys. Rev.} {\bf
  D92} (2015) 011501}, [\href{http://arxiv.org/abs/1504.08359}{{\tt
  1504.08359}}].

\bibitem{Sexty:2013ica}
D.~Sexty, \emph{{Simulating full QCD at nonzero density using the complex
  Langevin equation}},
  \href{http://dx.doi.org/10.1016/j.physletb.2014.01.019}{\emph{Phys. Lett.}
  {\bf B729} (2014) 108--111}, [\href{http://arxiv.org/abs/1307.7748}{{\tt
  1307.7748}}].

\bibitem{Fodor:2015doa}
Z.~Fodor, S.~D. Katz, D.~Sexty and C.~Török, \emph{{Complex Langevin dynamics
  for dynamical QCD at nonzero chemical potential: A comparison with
  multiparameter reweighting}},
  \href{http://dx.doi.org/10.1103/PhysRevD.92.094516}{\emph{Phys. Rev.} {\bf
  D92} (2015) 094516}, [\href{http://arxiv.org/abs/1508.05260}{{\tt
  1508.05260}}].

\bibitem{Aarts:2014kja}
G.~Aarts, F.~Attanasio, B.~Jäger, E.~Seiler, D.~Sexty and I.-O. Stamatescu,
  \emph{{Exploring the phase diagram of QCD with complex Langevin
  simulations}}, {\emph{PoS} {\bf LATTICE2014} (2014) 200},
  [\href{http://arxiv.org/abs/1411.2632}{{\tt 1411.2632}}].

\bibitem{Aarts:2014fsa}
G.~Aarts, F.~Attanasio, B.~Jäger, E.~Seiler, D.~Sexty and I.-O. Stamatescu,
  \emph{{QCD at nonzero chemical potential: recent progress on the lattice}},
  \href{http://dx.doi.org/10.1063/1.4938590}{\emph{AIP Conf. Proc.} {\bf 1701}
  (2016) 020001}, [\href{http://arxiv.org/abs/1412.0847}{{\tt 1412.0847}}].

\bibitem{Aarts:2015yba}
G.~Aarts, F.~Attanasio, B.~J{\"a}ger, E.~Seiler, D.~Sexty and I.-O. Stamatescu,
  \emph{{The phase diagram of heavy dense QCD with complex Langevin
  simulations}},
  \href{http://dx.doi.org/10.5506/APhysPolBSupp.8.405}{\emph{Acta Phys. Polon.
  Supp.} {\bf 8} (2015) 405}, [\href{http://arxiv.org/abs/1506.02547}{{\tt
  1506.02547}}].

\bibitem{Aarts:2015yuz}
G.~Aarts, F.~Attanasio, B.~J{\"a}ger, E.~Seiler, D.~Sexty and I.-O. Stamatescu,
  \emph{{Insights into the heavy dense QCD phase diagram using Complex Langevin
  simulations}},  in \emph{{Proceedings, 33rd International Symposium on
  Lattice Field Theory (Lattice 2015)}}, 2015.
\newblock \href{http://arxiv.org/abs/1510.09100}{{\tt 1510.09100}}.

\bibitem{Aarts:2015tyj}
G.~Aarts, \emph{{Introductory lectures on lattice QCD at nonzero baryon
  number}}, \href{http://dx.doi.org/10.1088/1742-6596/706/2/022004}{\emph{J.
  Phys. Conf. Ser.} {\bf 706} (2016) 022004},
  [\href{http://arxiv.org/abs/1512.05145}{{\tt 1512.05145}}].

\bibitem{Parisi:1980ys}
G.~Parisi and Y.-s. Wu, \emph{{Perturbation Theory Without Gauge Fixing}},
  {\emph{Sci. Sin.} {\bf 24} (1981) 483}.

\bibitem{Damgaard:1987rr}
P.~H. Damgaard and H.~H{\"u}ffel, \emph{{Stochastic Quantization}},
  \href{http://dx.doi.org/10.1016/0370-1573(87)90144-X}{\emph{Phys. Rept.} {\bf
  152} (1987) 227}.

\bibitem{Borsanyi:2012zs}
S.~Bors\'anyi et~al., \emph{{High-precision scale setting in lattice QCD}},
  \href{http://dx.doi.org/10.1007/JHEP09(2012)010}{\emph{JHEP} {\bf 09} (2012)
  010}, [\href{http://arxiv.org/abs/1203.4469}{{\tt 1203.4469}}].

\bibitem{link}
{The numerical data for the observables (density, Polyakov look, their
  susceptibilities and Binder cumulants), including statistical uncertainty,
  can be obtained from the source file of the ArXiv submission,
  \url{https://arxiv.org/abs/1606.05561}}.

\bibitem{Binder:1981sa}
K.~Binder, \emph{{Finite size scaling analysis of Ising model block
  distribution functions}},
  \href{http://dx.doi.org/10.1007/BF01293604}{\emph{Z. Phys.} {\bf B43} (1981)
  119--140}.

\bibitem{deForcrand:2002hgr}
P.~de~Forcrand and O.~Philipsen, \emph{{The QCD phase diagram for small
  densities from imaginary chemical potential}},
  \href{http://dx.doi.org/10.1016/S0550-3213(02)00626-0}{\emph{Nucl. Phys.}
  {\bf B642} (2002) 290--306},
  [\href{http://arxiv.org/abs/hep-lat/0205016}{{\tt hep-lat/0205016}}].

\bibitem{Wolff:2003sm}
{\scshape ALPHA} collaboration, U.~Wolff, \emph{{Monte Carlo errors with less
  errors}}, \href{http://dx.doi.org/10.1016/S0010-4655(03)00467-3,
  10.1016/j.cpc.2006.12.001}{\emph{Comput. Phys. Commun.} {\bf 156} (2004)
  143--153}, [\href{http://arxiv.org/abs/hep-lat/0306017}{{\tt
  hep-lat/0306017}}].

\bibitem{juqueen}
{J{\"u}lich Supercomputing Centre}, \emph{{JUQUEEN: IBM Blue Gene/Q
  Supercomputer System at the J{\"u}lich Supercomputing Centre}},
  \href{http://dx.doi.org/10.17815/jlsrf-1-18}{\emph{Journal of large-scale
  research facilities} {\bf 1, A1} (2015) }.

\end{thebibliography}\endgroup
\bibliographystyle{JHEP}

\end{document}